%% file: QCMC.tex
\documentclass[aps,pra,onecolumn]{revtex4}
\usepackage{epsfig,amssymb,amsmath,mathrsfs,amsthm,graphicx,float,color}
\usepackage{subfigure}
\usepackage[matrix,frame,arrow]{xypic}\input{myQcircuit}
\newcommand{\N}{\mathbb{N}}

\newcommand{\R}{\mathbb{R}}
\newcommand{\C}{\mathbb{C}}

\newcommand{\set}[1]{\mathsf{#1}}
\newcommand{\grp}[1]{\mathsf{#1}}
\newcommand{\spc}[1]{\mathcal{#1}}

\def\d{{\rm d}}

\def\>{\rangle}
\def\<{\langle}

\newcommand{\st}[1]{\mathbf{#1}}
\newcommand{\bs}[1]{\boldsymbol{#1}}     % allows you to make boldface Greek letters  

\newcommand{\map}[1]{\mathcal{#1}}
\newcommand{\Tr}{\operatorname{Tr}}

\newtheorem{theo}{Theorem}
\newtheorem{ax}{Axiom}
\newtheorem{lemma}{Lemma}

\newtheorem{protocol}{Protocol}

\begin{document}
\title{Quantum superreplication of states and gates}
\author{Giulio Chiribella\footnote{Corresponding author: giulio@cs.hku.hk} and Yuxiang Yang}
\affiliation{Department of Computer Science, The University of Hong Kong, Pokfulam Road, Hong Kong}
\begin{abstract}
While the no-cloning theorem forbids the perfect replication of quantum information, it is sometimes possible to produce    large numbers  of replicas  with  vanishingly small error.    This phenomenon,  known as quantum superreplication, can take place both for quantum states and quantum gates.  
 The aim of this paper is to  review the central features of  quantum superreplication, providing a unified view on the existing results.    
 % and emphasizing the connections between superreplication and other tasks, such as  programming gates and using them as a resource for state preparation.  
 The paper also includes new results. In particular,  we show that, when quantum superreplication can be achieved, it can be achieved through estimation, up to an error of size  $O(M/N^2)$, where $N$ and $M$ are the number of input and output copies, respectively.  Quantum strategies still offer an advantage for superreplication, in that they allow for an exponentially faster reduction of the error. 
 Using the relation with estimation, we provide \emph{i)} an alternative proof of the optimality of the Heisenberg scaling of quantum metrology, \emph{ii)} a strategy to estimate arbitrary unitary gates with mean square error scaling  as $\log N/N^{2}$, and \emph{iii)}   a protocol that generates $  O(N^2)$ nearly perfect copies of a generic pure state  $U|0\>$  while using the corresponding gate $U$ only  $N$ times.    Finally, we   point out that superreplication can be achieved using interactions among $k$ systems, provided that $k$ is large compared to $M^2/N^{2}$. 
%[  ADD A COMMENT ON THE FACT THAT OUR LIMITS DO NOT HOLD IN INFINITE DIMENSION, ADD COMMENT ON COHERENT STATES]
\end{abstract}

\maketitle

\section{Introduction}

The no-cloning theorem \cite{wootters-zurek-1982-nature,dieks-1982-pla} is one of the cornerstones of quantum information theory, with implications permeating the whole field \cite{scarani-iblisdir-2005-rmp,cerf-fiurasek-2006-pio}.  
 %The impossibility to copy non-orthogonal quantum states played  a pivotal role in the development of   quantum  protocols, 
 Most famously, the impossibility to copy non-orthogonal quantum states provides a working principle for quantum cryptography   with applications to key distribution \cite{bennett-brassard-1984,ekert-1991-prl}, unforgeable banknotes \cite{wiesner-1983-ACM}, and secret sharing \cite{hillery-buzek-berthiaume-1999-pra}.    
 The no-cloning theorem  forbids perfect cloning.  A natural question, originally asked by Bu\v zek and Hillery \cite{buzek-hillery-1996-pra},  is how well cloning can be approximated by the processes allowed by quantum mechanics.   The question is relevant both to cryptographic applications  and to the foundations of quantum theory,  shedding light on  relation between quantum    and classical copy machines  \cite{gisin-massar-1997-prl, bruss-ekert-macchiavello-1998-prl, bae-acin-2006-prl,chiribella-dariano-2006-prl, chiribella-2010-tqc, chiribella-yang-2014-njp} and providing benchmarks  that certify the advantages of quantum information processing over classical information processing \cite{braunstein-fuchs-2000-jmo,hammerer-wolf-2005-prl,adesso-chiribella-2008-prl,chiribella-xie-2013-prl,chiribella-adesso-2014-prl}. 
 
 Among the  cloning machines allowed by quantum mechanics, one can distinguish two types:   deterministic  and probabilistic machines.  Deterministic machines produce approximate copies with certainty, while probabilistic machines sometimes produce a failure message  indicating that the copying process has gone wrong. In general,  probabilistic machines can produce more accurate copies at the price of a reduced   probability of success. For example,  Duan and Guo \cite{duan-guo-1998-prl} showed that a set of  non-orthogonal states can be cloned without error as long as they  are linearly independent.   More recently, Fiur\'a\v sek  \cite{fiurasek-2004-pra} showed that  probabilistic cloners can  offer an advantage even for linearly dependent states, including~e.~g.~coherent states of  harmonic oscillators  with known amplitude.  Ralph and Lund \cite{ralph-lund-2009-QCMC} proposed a concrete optical setup achieving  noiseless probabilistic amplification and cloning of coherent states. 
 The possibility of noiseless probabilistic amplification was later extended to the case where the coherent state amplitude is Gaussian-distributed \cite{chiribella-xie-2013-prl}.  Although the probability of success  vanishes as the accuracy increases, a nearly perfect amplification of coherent states has been observed experimentally for small values of the amplitude  \cite{xiang-ralph-2010-natphoton,ferreyrol-barbieri-2010-prl,usuga-muller-2010-natphys,zavatta-fiurasek-2011-natphoton}.  
%parameter estimation \cite{fiurasek-2006-njp,gendra-ronco-2013-prl,gendra-ronco-2013-pra}, 

The recent developments on probabilistic amplification and cloning of coherent states motivate  the search for new scenarios where cloning of non-orthogonal states can be achieved with vanishingly small error.  In \cite{chiribella-yang-2013-natcomm} we considered the  task of cloning \emph{clock states}, that is,  quantum states generated by time evolution under a known Hamiltonian.   Here, the cloner is given $N$ identical copies of the same clock state and attempts  to produce $M  =   M(N)$ copies. Inspired by the asymptotic framework of information theory, we considered the scenario where $N$ is large.  Here, the main question is how fast $M (N)$ can grow  under the   condition  that the copying process is \emph{reliable}, meaning that the  global error  on all the  output copies vanishes in the large $N$ limit. 
 For deterministic cloners,  we showed that the number of reliable copies has to scale as 
  \begin{align}\label{noclon}
 M  (N)=  N \,  [ 1  +  f(N)]  \qquad  {\rm with}  \qquad  \lim_{N\to \infty}  f (N)  = 0  \, .     
 \end{align}
 In words, this means that the number of extra copies produced by the cloner must be negligible compared to the number of input copies.    This result can be seen as an \emph{asymptotic no-cloning theorem}, which   extends the original no-cloning theorem from perfect cloners to cloners that become perfect in the asymptotic limit.  
The origin of the asymptotic no-cloning theorem is the requirement  that the error vanishes \emph{globally} on all the clones.      In general,  using tomography it is  easy to produce copies that, individually, have an error vanishing order $1/N$. However, this does not guarantee that the  error  vanishes at the global level.  Indeed,  the errors accumulate when multiple copies are examined jointly.   
 When this fact is taken into account, it turns out that  only a cloner satisfying Eq. (\ref{noclon}) can have a  vanishing error at the global level.   For probabilistic cloners of clock states, the situation is  different:  %  we found out that the asymptotic no-cloning theorem can be evaded:
  the number of reliable copies produced by a probabilistic machine can scale as 
  \begin{align}\label{ratelim}
M(N)    =  N     [ 1  +  f(N)]  \qquad  {\rm with}  \qquad   f (N)  =   \Theta  \left( N^{\delta}\right) \,  ,   \delta <1  \, .   
   \end{align}
%  where  the origin of the upper bound $\delta  <1$  is   the Heisenberg limit of quantum metrology \cite{metro science paper}.   
  When $\delta$ is zero, this means that the number of extra copies scales as $N$,  evading  the limitation posed by the  asymptotic no-cloning theorem. When $\delta$ is larger than zero,  the number of extra copies grows faster than $N$ and one can produce up to  $M  =  \Theta  ( N^{2})$ copies. 
   In both cases,  we refer to  this  phenomenon as  \emph{superreplication}, emphasizing the fact that  the number of extra-copies grows beyond the limits imposed by the asymptotic no-cloning theorem.    Note however, that the number of output copies allowed by Eq. (\ref{ratelim}) can grow at most at rate
   $M  =  \Theta \left(N^{2-\epsilon}\right)$, where $\epsilon$ is an arbitrarily small constant. It turns out that, in finite dimensions, the quadratic replication rate is the ultimate limit for the  superreplication of clock states, in close connection with  the Heisenberg limit of quantum metrology \cite{giovannetti-lloyd-2004-science,chiribella-yang-2013-natcomm}.  
   
In addition to producing a large number of extra copies, probabilistic cloners exhibit a better scaling of the error, which vanishes as $\exp  [-c  N^2/M  (N)]$ for a suitable constant $c>0$ depending on the particular  set of clock states under consideration.  In contrast, the error for deterministic machines  can scale at most as $1/N^4$  \cite{chiribella-yang-2013-natcomm}.  
 The benefits  of probabilistic cloners, however, do not come for free. The price to pay is a very small probability of success: precisely,  the probability of superreplication has to be small compared to $\exp  \left[-  M/N \right]$. This fact poses a severe limitation on the ability to observe superreplication, especially when $M$ grows much faster than $N$.   Nevertheless, the cloner  can be devised in such a way that, when superreplication fails, the original input state is almost undisturbed. Precisely, Winter's gentle measurement lemma \cite{winter-1999-ieee} implies that  the  error introduced by the failure of superreplication   scales as $O\left(  \sqrt{  p_{\rm succ}}  \right)$,  where  $p_{\rm succ}$ is the probability of success. 
     For superreplication processes with $\delta  >  0$, this means that the error introduced by the failure of superreplication vanishes faster than  $\exp  \left[-  N^{\delta/2} \right]$.   As a consequence,  the state resulting from a failed superreplication  can still be used to achieve standard, deterministic cloning with asymptotically optimal performance: for large $N$,  the error $\exp  \left[-  N^{\delta/2} \right]$  is covered by the  error introduced by deterministic cloning.     In summary, superreplication is a rare event, but trying to observe it  does not   prevent the application of standard deterministic cloning techniques.  

Interestingly, the limitation on the probability of success  is lifted if we consider the problem of cloning gates, instead of states.  Cloning a quantum gate  means simulating $M$ uses of it while actually using it only $N<M$ times.   A no-cloning theorem for quantum gates was proven by D'Ariano, Perinotti, and one of the authors \cite{chiribella-dariano-2008-prl}, who showed  that no quantum protocol can perfectly clone a generic quantum gate. 
   Recently, D\"ur, Sekatski, and Skotiniotis \cite{dur-sekatski-2015-prl} analyzed  the cloning of   phase gates,~i.~e.~unitary gates describing the time evolution with a known Hamiltonian. They devised   a quantum network that simulates up to  $N^2$ uses of an unknown phase gate by using it only $N$ times. 
 The network works deterministically and has vanishing error on average over all input  states.   We refer to this phenomenon as \emph{gate superreplication}. 
 The discovery by  D\"ur \emph{et al} opened the question whether superreplication  can be achieved not only for phase gates, but also for arbitrary quantum gates.  In Ref. \cite{chiribella-yang-2015-prl} we answered the question in the affirmative, constructing a universal quantum network that  replicates completely unknown unitary gates.  The network has vanishing error on almost all input states, except for a vanishingly small fraction of the Hilbert space. 
   These ``bad states" can be characterized explicitly,  making it easy to identify the applications where gate superreplication can be safely employed.  
  
In this paper we review the key facts about superreplication of states and gates, unifying the  ideas presented in the literature and emphasizing the  connections between superreplication and other tasks, such as estimating quantum gates and generating states using gates as oracles.   
 In addition to the review part, the paper contains a number of new results:  
 \begin{enumerate}
 \item We show that, whenever  quantum superreplication is achievable, it can  also be achieved through estimation. However, the error  for estimation-based strategies will vanish with a power law, while  the error for the genuine quantum strategies vanishes faster than any polynomial.   
 \item We establish an equivalence between  optimality of the Heisenberg scaling in quantum metrology and the ultimate limit on the rate of superreplication, encapsulated in Eq. (\ref{ratelim}). On the one hand,  the Heisenberg scaling implies the impossibility to produce more than $M =  O(N^2)$ copies of a clock state with  vanishing error \cite{chiribella-yang-2013-natcomm}.  On the other hand, here we prove the converse result, showing that  the  limit on the replication rate set by Eq. (\ref{ratelim}) implies the optimality of the Heisenberg scaling.  
     \item We explore  the possibility to achieve superreplication without acting globally on all the $N$ input copies and on  $M-N$ additional  blank copies.     Specifically, we consider strategies where the $N$ input copies are divided into subgroups and each subgroup generates new copies, mimicking a   scenario where  groups of cells fuse together and then split into approximate clones.   For replication strategies of this form, we show that interactions among groups of $O(M^2/N^{2-\epsilon})$  particles are necessary and sufficient to achieve superreplication.   
     With respect to the original superreplication protocol, the modified protocol offers a reduction of the size of the interactions by a factor $ M/N^{2-\epsilon}$, which is asymptotically large for all the replication rates allowed by Eq. (\ref{ratelim}). 
     \item We construct a simple  protocol that estimates  an unknown unitary gate with mean square error $\log N/N^2$.   The protocol uses gate superreplication to produce $M=  \Theta  (N^2/\log N)$ copies and gate tomography to estimate the gate within a mean square error of size $ 1/M$.  The resulting scaling of the error  is close to the optimal scaling $1/N^2$  \cite{chiribella-dariano-2004-prl,bagan-baig-2004-pra,hayashi-2006-pla,kahn-2007-pra}. 
       \item We  analyze the task of  generating $M$  copies of a quantum state $U| 0\>$ given $N$ uses of a completely unknown unitary gate $U$  \cite{chiribella-yang-2015-prl}.   In this setting we show that $M  =  \Theta (N^{2-\epsilon})$ copies of the state can be generated with fidelity going to 1 in the large $N$  limit, for every $\epsilon>0$.  
 \item  We examine the replication of quantum gates in the worst case over all possible input states. In the worst-case  scenario, we establish an \emph{asymptotic no-cloning theorem for quantum gates}, stating that the number reliable copies of the gate  scales as  $M(N)   =  N   [1  +  f(N)]$, where $f(N)$ vanishes in the large $N$ limit.  
   \end{enumerate}

 The paper is organised as follows. Section \ref{sec:state} introduces the task of asymptotic cloning and precisely analyses the phenomenon of quantum state superreplication. Section \ref{sec:compare} analyses the relationship between state superreplication and quantum metrology.  Section \ref{sec:network} discusses the scale of the  interactions needed for state superreplication. In Section \ref{sec:gate} we review the existing results on the superreplication of quantum gates. 
  Several applications of gate superreplication are presented in Section  \ref{sec:application}.  Finally,  the conclusions are drawn  in Section \ref{sec:conclusion}. 

\section{Quantum state superreplication}\label{sec:state}

%\subsection{Quantum cloning and the asymptotic no-cloning theorem}

\noindent{\bf Deterministic  cloners.} Although perfect cloning of  non-orthogonal quantum states is forbidden by the no-cloning theorem \cite{wootters-zurek-1982-nature,dieks-1982-pla}, approximate cloning can be realized to various degrees of accuracy, depending on the set of states to be cloned   \cite{gisin-massar-1997-prl, bruss-ekert-macchiavello-1998-prl, bae-acin-2006-prl,chiribella-dariano-2006-prl, chiribella-2010-tqc, chiribella-yang-2014-njp}. Consider a scenario  where one is given  $N$ identical systems, each of them  prepared in the same  $|\psi_x\>$, and the goal is to generate $M$ systems in a state close to $M$ perfect copies of the state $|\psi_x\>$.   For the moment, we assume the cloning process to be deterministic, meaning that it produces approximate clones with unit probability.  
Mathematically, a deterministic cloner is described by a completely positive trace-preserving linear map  $\map C$, a.k.a. a quantum channel, sending states on the Hilbert space $\spc H^{\otimes N}$ to states on the Hilbert space $\spc H^{\otimes M}$, where $\spc H$ is the Hilbert space of a single copy.  The cloning accuracy can be quantified by the worst case fidelity between the ideal state of $M$ perfect copies and  the actual output state of the cloner. Explicitly, the average fidelity is given by 
\begin{align}\label{Fdet}
F_{\rm det}  [N\to M]=\sum_x p_x \Tr\left[|\psi_x\>\<\psi_x|^{\otimes M}\map{C}\left(|\psi_x\>\<\psi_x|^{\otimes N}\right)\right] \, ,
\end{align}
where $p_x$ is the prior probability of the state $|\psi_x\>$.  The quantum channel that maximizes the fidelity is the optimal $N$-to-$M$ cloner for the ensemble   $\{  |\psi_x\> \, , p_x\}$.   In general, the form of the optimal channel varies for different ensembles, depending both on the states and on their prior probabilities.      

\medskip
\noindent{\bf Example: the equatorial qubit cloner.}  The general settings of optimal cloning can be nicely illustrated in the example of equatorial qubit states  \cite{fan-matsumoto-2001-pra,bruss-cinchetti-2000-pra, dariano-macchiavello-2003-pra}.    Here, the ensemble consists of states of the form $|\psi_t \>=(|0\>+e^{-i t}|1\>)/ \sqrt{2}$ where $\theta$ is drawn uniformly at random from the interval  $[0,2\pi)$. The input state can be expanded as
\begin{align*}
|\psi_t\>^{\otimes N}=\frac{1}{2^{N/2}}\sum_{n=0}^{N}\sqrt{{N\choose n}}e^{- i  n  t }|N,n\>
\end{align*}
where $\{|N,n\>~|~n=  0  , \dots, N\}$  is the orthonormal basis of the Dicke states, defined as   
\[|N,n\>:=\frac1{\sqrt{N! \, n! \,(N-n)! }}  \, \sum_{\pi  \in  \grp S_N}  \,  U_{\pi}  \, |0\>^{\otimes (N-n)}  \,  |1\>^{\otimes n}  \, ,\]  
 $\pi$ being a permutation of the $N$ qubits, $U_{\pi}$ being the unitary operator that implements the permutation $\pi$, and the sum running over the symmetric group $\grp S_N$.  
   
For convenience of discussion, we  focus on the case where  both $N$ and $M$ are even. In this case, the optimal cloning channel   has the  simple  form $\map{C}(\rho)=V\rho V^\dag$, where $V$ is the isometry  defined as  \cite{dariano-macchiavello-2003-pra}
\begin{align*}
V \left |N,  \frac N 2+m  \right\>=\left |M,  \frac M 2+m \right\> \, , \qquad   m\in\left[-\frac N2, \frac N2\right]  \, .
\end{align*}
Plugging the above relation into the definition of the average fidelity  (\ref{Fdet}), one obtains the optimal value
\begin{align}
\nonumber F_{\rm det}  [N\to M] &=\frac{1}{2^{N+M}}\left[\sum_{m=-N/2}^{N/2}\sqrt{{N\choose\frac N2+m}{M\choose \frac M2+m}}\right]^2\\
\label{Fdet-equatorial}
 &   \approx  \frac{2\sqrt{MN}}{M+N}   \qquad     N,  M  \gg 1  \, .
\end{align} 

\medskip
\noindent{\bf   The asymptotic no-cloning theorem.}   Inspired by the asymptotic framework of information theory, we  now focus on quantum cloners in the limit $N\to \infty$.   We model the cloning process as a sequence of cloners $ \left(   \map C_N  \right )_{N  \in  \N}$, where $\map C_N$ transforms  $N$  copies into $M(N)$ approximate copies for a given function $M(N)$.  
    We call  the  cloning process \emph{reliable}, if the cloning fidelity goes to one in the asymptotic limit, namely 
    \[  \lim_{N\to\infty} F_{\rm det}  [N\to M(N)]=1 \,.
    \] 
    The key question here is how many extra copies can be produced reliably.   Can we produce $N$ extra copies or more?     To provide a rigorous answer, we define the \emph{rate} of a cloning process as  
    \[   \alpha   :=    \liminf_{N\to\infty}    \frac{  \log  [M(N)  -N  ]}{\log N}  \, . \]  
  In words: if the rate is $\alpha$, then the number of output copies grows as $M(N)  =    N  +  \Theta  ( N^{\alpha})$.    We say that a cloning rate  $\alpha$ is \emph{achievable} iff there exists a reliable cloning process  that has  rate equal to $\alpha$.  
   For a given set of states, the  task  is  to find the maximum achievable rate over all cloning processes.      
   In the case of deterministic processes,  the following theorem  places a tight limit on the number of copies that can be produced reliably:

\begin{theo}[Asymptotic no-cloning theorem for quantum states  \cite{chiribella-yang-2013-natcomm}]\label{thm:no-cloning}
No deterministic process can reliably clone a continuous set of quantum states  at a rate $\alpha \ge 1$. 
\end{theo}
The proof idea comes from the standard quantum limit of metrology \cite{giovannetti-lloyd-2004-science} which limits the precision in the deterministic estimation of a parameter $t$ encoded into   $N$ product copies of a  state $|\psi_t\>$.   The standard quantum limit states that the mean square error   vanishes as  $c/N$, where $c$ is a constant that depends on the encoding $t \mapsto  |\psi_t\>$.     Intuitively, a deterministic cloner violating theorem \ref{thm:no-cloning} would  contradict the standard quantum limit, because one could increase the precision by first cloning the probe states and then measuring them.  

The argument based on the standard quantum limit of metrology is partly heuristic.  A complete argument can be made for \emph{clock states},~i.~e.~quantum states of the form 
 \[  |\psi_t\>=e^{-itH}|\psi\>\, , \qquad t\in\R \, ,    H^\dag =  H \, .\]   
The argument is conceptually independent of the standard quantum limit and allows one to prove a strong converse of the asymptotic no-cloning theorem: 
\begin{theo}[Strong converse of the asymptotic no-cloning theorem \cite{chiribella-yang-2013-natcomm}]\label{theo:strong1}
Every deterministic process that clones clock states at a rate $\alpha   >1$ will have necessarily a vanishing fidelity  in the large $N$ limit.  
\end{theo}

The asymptotic no-cloning theorem implies that a deterministic cloning process  cannot produce more than a negligible number of extra replicas in the asymptotic limit. 
 Let us check it a few examples. The first example is the cloning of equatorial qubit states, introduced earlier in the paper.    In the large $N$ limit, the cloning fidelity  is given by $  F   \approx 2\sqrt{MN}/(M+N)$  [cf. Eq. (\ref{Fdet-equatorial})]. It is straightforward to see that any   achievable cloning rate must be smaller than one. 
 Another example is the universal cloning of pure states.   For $d$-dimensional quantum systems, the optimal fidelity is $F ={d+N-1\choose N}/{d+M-1\choose M}$ \cite{werner-1998-pra}.  For large $N$ and $M$, the fidelity  scales as $(N/M)^{d-1}$   and converges  to one if and only if the cloning rate satisfies the condition  $\alpha<1$.   The example of universal cloning shows that one can always find a deterministic process that reliably produces  $M =    N +  \Theta  (N^\delta  ) $ copies, for every desired exponent $\delta >  0$.   
 
  It is important to stress that the asymptotic no-cloning theorem holds for \emph{continuous} sets of states, but does not  place any restriction on the cloning rate on discrete  sets of states.  For example, every finite set of quantum states can be cloned with vanishing error in the large  $N$  limit, no matter how large  is $M$   \cite{chiribella-yang-2014-njp}.   Hence, for finite sets of states every rate is achievable.

%\subsection{Quantum state superreplication and the Heisenberg limit}

\medskip
\noindent{\bf Probabilistic cloners.} The asymptotic no-cloning theorem poses a stringent limit on the ability to replicate information.   
     In the following we  will see that the limit can be broken by allowing the cloner to be probabilistic. 
 % In this case, rates up to we show that the cloning rate jumps up to be quadratic.

To analyze  probabilistic  cloners it is useful to introduce the notion of  quantum instrument \cite{davies-lewis-1970-cmp, ozawa-1984-jmp}.   A quantum instrument consists of an indexed set of  completely positive, trace non-increasing maps  (a.k.a. quantum operations) $\{\map{M}_1,\map{M}_2,\dots\}$. 
 When an input state $\rho$ is fed into the quantum instrument, the quantum operation $\map{M}_i$ occurs with probability $p_i=\Tr[\map{M}_i(\rho)]$ and the instrument outputs a quantum system in the state $\rho'  =\map{M}_i(\rho)/p_i$. In the context of our cloning problem, we consider a quantum instrument consisting of two quantum operations $\map{M}_{\rm yes}$ and $\map{M}_{\rm no}$, with  $\map M_{\rm yes}$ describing the realization of a successful cloning process.  When acting on the $N$-copy input state $|\psi_x\>^{\otimes  N}$, the probabilistic cloner succeeds with probability 
\begin{align} \label{px}
p  ({\rm yes } |  x)=\Tr\left[\map{M}_{\rm yes}(|\psi_x\>\<\psi_x|^{\otimes N})\right] \, ,
\end{align} 
in which case it   produces the $M$-copy output state
\begin{align}\label{rhox}
\rho_x'  =  \map{M}_{\rm yes}(|\psi_x\>\<\psi_x|^{\otimes N})/p  (  {\rm yes}|  x) \, .
\end{align}   
%Otherwise, the operation $\map{M}_{\rm no}$ is performed and its output is discarded. 

Conditionally on the success of the cloning process, the average fidelity is equal to 
\begin{align}\label{Fbayes}
F_{\rm prob}  [N\to M]&    = \sum_x  \,   p(  x  |  {\rm yes})  \,   \Tr \left[   |\psi_x\>\<\psi_x|^{\otimes M}    \,  \rho_x' \right] \,  ,
\end{align}
where $p(  x|  {\rm  yes})$ is the conditional  probability given by Bayes' rule. Inserting Eqs. (\ref{px}) and (\ref{rhox}) in the above expression, we obtain the explicit formula
\begin{align} 
\label{Fprob}
 F_{\rm prob}  [N\to M] =~\frac{\sum_{x}p_x  \Tr\left[|\psi_x\>\<\psi_x|^{\otimes M}\map{M}_{\rm yes}(|\psi_x\>\<\psi_x|^{\otimes N})\right]}{\sum_{y}p_y \, \Tr\left[\map{M}_{\rm yes}(|\psi_y\>\<\psi_y|^{\otimes N})\right]} \, .
\end{align}
The maximum  fidelity over all possible quantum instruments represents the  ultimate performance allowed  by  quantum mechanics,  even if  the probability of success is arbitrarily small. In the following we will focus on the maximum fidelity in the asymptotic limit of large $N$ and $M$.  

\medskip
\noindent{\bf Superreplication of equatorial qubit states.} Let us start from the simple example of   equatorial qubit states.    In this case, the optimal  probabilistic cloner is given by a quantum instrument with  successful operation $\map{M}_{\rm yes}(\rho)=Q\rho Q^\dag$ given by 
\begin{align}\label{cloner}
Q \, \left |N,  \frac N 2  + m \right\>=\sqrt{\frac{{M\choose \frac{M}2+m}}{{M\choose\frac{M+N}2}{N\choose \frac{N}2+m}}}   \left |M,\frac M 2 + m \right\> \, , \qquad m\in  \left[-\frac N2,  \frac N2  \right]
\end{align}
(recall that we are assuming $N$ and $M$ even for simplicity of presentation).  
Inserting the expression of the optimal cloner in the the fidelity  (\ref{Fprob}), we obtain the optimal value
\begin{align}
F_{\rm prob}  [N\to M]&=\frac{1}{2^{M}}\sum_{m=-N/2}^{N/2}{M\choose\frac{M}2+m}\nonumber\\
&\ge 1-2\exp\left(-\frac{N^2}{2M}\right),\label{Fclock}
\end{align}
having used Hoeffding's inequality.   One can immediately see that any cloning rate $\alpha<2$ is achievable: in this case, the cloning error vanishes with $N$ faster than the inverse of every polynomial.   Interestingly, such rapid decay of the error is impossible for deterministic cloners, whose error can vanish at most as  $1/N^4$  \cite{chiribella-yang-2013-natcomm}.   In summary, allowing the cloner to claim failure results into two advantages: \emph{i)}  the cloning rate can be boosted beyond the limits of the asymptotic no-cloning theorem, and \emph{ii)} the cloning error vanishes with $N$ at a speed that could not be achieved by deterministic cloners.   The contrast between deterministic and probabilistic cloners can be also  seen in the non-asymptotic setting: for example,  when $N=20$ and $M=120$, the probabilistic fidelity maintains the high value $94.52\%$, while the deterministic fidelity drops to $69.39\%$. 

The optimal probabilistic cloner dramatically outperforms all deterministic cloners in terms of   rate and accuracy. However, these advantages do not come for free: the price to pay is that the  probability of success vanishes in the large $N$ limit.  For example,  the probability of success for   qubit cloner of Eq. (\ref{cloner}) is given by  
\begin{align*}
p_{\rm yes}  [N\to M]  &=   \frac{   \left|  \<\psi_t  |^{\otimes  M}   Q   | \psi_t\>^{\otimes N}    \right|^2  }{     \left|  \<\psi_t  |^{\otimes  N}  Q^\dag Q   | \psi_t\>^{\otimes N }   \right|^2}  \\
&   =   \frac{1}{2^{N}{M\choose\frac{N+M}2}}\sum_{m=-N/2}^{N/2}{M\choose\frac{M}2+m} \, , \qquad \forall t\in [0, 2\pi) \, .
\end{align*}
  Using Stirling's approximation  one can see that the above  probability satisfies the asymptotic equality 
$$ 
p_{\rm yes}  [N\to M] \approx e^{-N \, \left(\ln2  -\frac{N}{M}\right)}  \, , \qquad  N \ll M  \, .$$
%In \cite{chiribella-yang-2013-natcomm} it was shown that the success probability can be increased by adjusting the cloner, yet the exponential decay is inevitable. 
 This example  indicates a trade-off between the cloning rate and the success probability. 
 In the following we will see that such trade-off is a generic feature of  state superreplication.

\medskip
\noindent{\bf General case: superreplication of quantum clocks.} The superreplication of  equatorial qubit states can be  generalised to the broad family of clock states, of the form  $|\psi_t\>=e^{-itH}|\psi\>$ with $t\in\R$ and $H^\dag =  H$.   For every family of clock states, one can find  a probabilistic  $N$-to-$M$ cloner that achieves fidelity \cite{chiribella-yang-2013-natcomm}
$$F_{\rm prob}  [N\to M]\ge 1-2K\exp\left(-\frac{2p_{\min}^2 N^2}{M}+\frac{4N}{MK}\right)$$
independently of $t$.    Here, $K$ is the number of distinct energy levels of the Hamiltonian $H$ and $p_{\min}$ is the smallest non-zero probability in the probability distribution of the energy for the state $|\psi\>$.  
 It is immediate to see  that every  cloning rate $\alpha<2$ is achievable, and thus clock states can be superreplicated. Again, the cloning error vanishes with $N$ faster than the inverse of every polynomial---a fast scaling that could not be achieved by deterministic cloners.   In other words, the probabilistic cloner produces not only \emph{more} but also \emph{better} replicas.

 The exponential decay of the success probability is  a necessary condition for  superreplication of clock states:     indeed, every reliable superreplication process must have a success probability that is vanishingly small compared to $\exp\left[-M/N\right]$.   In other words, as soon as a cloning process violates the asymptotic no-cloning theorem,  the probability of the process must vanish in the large $N$ limit.  
   A strong converse result can also be proven  \cite{chiribella-yang-2013-natcomm}:
   \begin{theo} 
    Every cloning process with rate $\alpha\ge 1$ and success probability of order   $\exp\left[-M(N)/ N\right]$ or larger  will necessarily have vanishing fidelity. 
    \end{theo} 
  On the other hand, it is possible to show that superreplication \emph{can} be achieved with success probability  $p_{\rm yes}     > \exp\left[  -  M / N^{1-\epsilon}\right] $  for every $\epsilon  >0$.

Informally, we can think of superreplication as a lottery where one invests the initial  copies of the state in the hope to obtain a much larger number of approximate copies. Despite the low probability of success, the superreplication lottery has almost no risk of loss: indeed, it is possible to guarantee that, when replication fails, the cloner returns the $N$ input copies, up to an error that is vanishingly small compared to   $\exp\left[-M/2N\right]$.   This result follows from the gentle measurement lemma  \cite{winter-1999-ieee}.   
   Since the error is small, the input copies can still be used for deterministic cloning.  Recall that the cloning error of deterministic cloners vanishes at most as $1/N^4$. Hence,  the error introduced by the failed attempt at  superrplication is negligible compared to the cloning error.   
 In summary, one can achieve the same  asymptotic performances of the optimal deterministic cloning, while still allowing for the chance to observe the exotic superreplication phenomenon.   
   
\begin{figure}\label{recycle}
\centering
\includegraphics[width=0.7\linewidth]{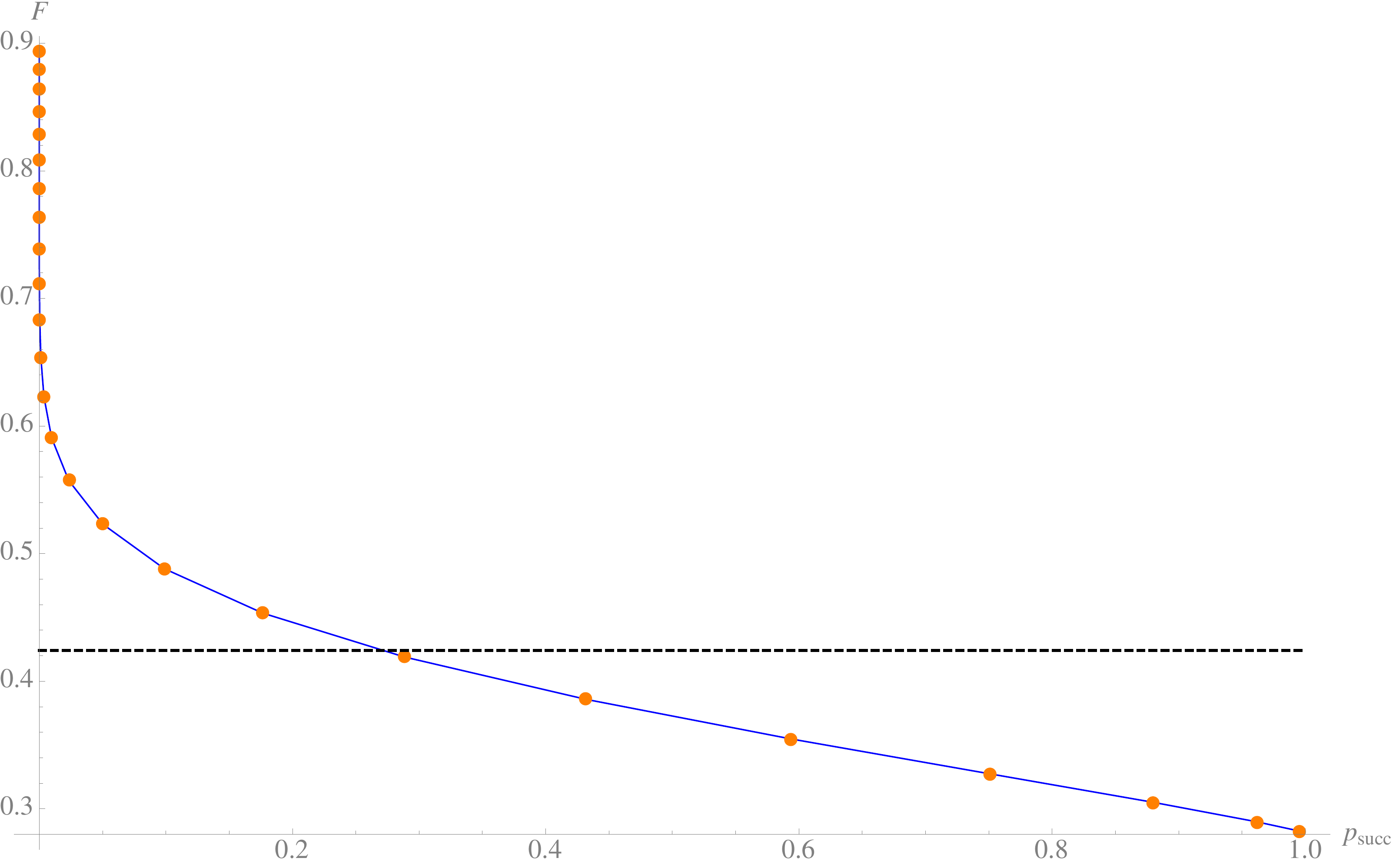}\\
\caption{{\bf Replication of equatorial qubit states through successive attempts of probabilistic cloning.} The figure shows the tradeoff between fidelity and success probability  for $N=50$ and $M=1000$. The points on the blue line represent successive attempts to generate $M$ copies through the optimal probabilistic process.    The black dashed line represents the fidelity of the optimal deterministic cloner.   In this example, the errors introduced by failed attempts cause the fidelity to fall below the optimal deterministic fidelity when the probability of success becomes larger than 20\%. }
\label{fig:qubit}
\end{figure}

Building on this observation, we can construct a cloner that makes a series of attempts to copy clock states probabilistically, with every failed attempt resulting into a small deterioration of the input copies  \cite{yang-chiribella-2015-arxiv}. This cloner increases the probability of success, at the price of a performance that decreases with the number of attempts.  In Figure \ref{recycle} we show a plot of the fidelity as a function of the total probability of success in the  case of equatorial qubit states with $N= 50$  and $M  =  1000$.   From the plot one can see the high fidelity/low probability region corresponding to the first attempts and the rapid decrease of the fidelity with the number of attempts. 

\medskip 

\noindent{\bf The Heisenberg limit for superreplication.}   We have seen that clock states can be superreplicated for every  cloning rate smaller than two.   Higher cloning rates are forbidden by the following 
\begin{theo}[Heisenberg limit for superreplication \cite{chiribella-yang-2013-natcomm}]\label{theo:HL}
In finite dimensions, no physical process can achieve a cloning rate larger than two for a  set of states containing clock states. 
\end{theo}
The origin of the limit on the cloning rate is  the Heisenberg limit of quantum  metrology \cite{giovannetti-lloyd-2004-science}, which can be extended to  probabilistic strategies in the case of finite dimensional systems \cite{chiribella-yang-2013-natcomm}.   The intuitive idea is the following: the Heisenberg limit  implies that  the probabilistic estimation of $t$ from the state $|\psi_t\>^{\otimes N}$ has a minimum mean square error scaling as  $N^{-2}$. If one could   produce  $M =\Theta (N^{\alpha})$  clones with sufficiently small error, then these clones could be used to estimate $t$.  Performing  individual measurements on the clones and collecting the statistics, one could make the mean square error as small as  as  $O(1/N^{\alpha})$.  Clearly, the Heisenberg limit implies  $\alpha\le 2$. 
 
The argument based on the Heisenberg limit is partially heuristic, because is relies on the assumption that the replication error vanishes sufficiently fast.    However, it is possible to make a complete argument based on the direct optimization of the probabilistic cloner. This argument is conceptually independent of the Heisenberg limit and allows to prove a strong converse  \cite{chiribella-yang-2013-natcomm}:   
\begin{theo}[Strong converse of the Heisenberg limit for superreplication]\label{theo:strong2}  Every physical process that replicates clock states with replication rate $\alpha  > 2$  will necessarily have vanishing fidelity, no matter how small is its probability of success.  
\end{theo}

Note however that  the restriction to finite-dimensional systems is essential for the above conclusions.   For infinite dimensional systems, the   probabilistic Heisenberg bound can easily become trivial.  For example, consider the set of all coherent states with fixed amplitude $r>0$, which can be seen as an infinite-dimensional example of clock states  
\[    |  \psi_t  \>   = e^{-i  t  \, a^\dag a } \,  |\psi_0\>  \, ,  \qquad     |\psi_0\>  =  e^{- r^2/2} \, \sum_{n=0}^\infty \frac {r^n}{n!} \,  |n\>  \, , \]       
with $a\,  |n\>  =  \sqrt{n} \,  |n-1\>$, $\forall n\in \N$.   Now, coherent states with known amplitude can be probabilistically cloned with fidelity arbitrarily close to 1  \cite{fiurasek-2004-pra}, by using Ralph and Lund's  noiseless probabilistic amplifier       \cite{ralph-lund-2009-QCMC}.    This means that we can pick every function $M(N)$  and build a probabilistic cloner that transforms $N$ input copies into $M(N)$ approximate copies with error smaller than, say,  $1/N$.    The cloning process constructed in this way is  reliable and can have \emph{every} desired rate  $\alpha$, thus breaching  the Heisenberg limit  $\alpha<2$.  The catch is, of course, that the probability of success will decay faster for processes with higher rate.   Pandey \emph{et al} provide a general bound \cite{pandey-jiang-2013-pra}, implying that a  replication process  with rate $\alpha$ must satisfy the relation
%ferrie-combes-2014-,
%chiribella-xie-2013-prl} 
$$p_{\rm yes}  [N\to M]\le \exp\left[-(\alpha-1)r^2 N^{\alpha}\ln N \right].$$

\noindent {\bf State superreplication beyond clock states.} 
 The superreplication of quantum clocks implies the superreplication of other relevant families of states, including,  the multiphase covariant states 
\[|\psi_{\bs \theta}\>=    \sqrt {p_0}  \,  |0\>  +  \sum_{j=1}^{d-1}\sqrt{p_j}e^{i\theta_j}| j\>   \qquad \theta_j  \in  [0,2\pi)  \, ,  \forall   j  = 1,\dots,  d-1  \, . \] 
To see that these states can be superreplicated, it is enough to consider the family of clock states     $  |\psi_t\>   =  e^{-it H}  \,   |\psi\>$, with $  |\psi\> =  \sum_{j=0}^{d-1}  \, \sqrt{p_j} \,   |j\>$ and $H  =  \sum_{j=0}^{d-1} \,   \sqrt{n_j}  \,   |j\>\<  j|$, where $n_j$ is the $j$-th prime number.    With this choice,  the closure of the set $\{  |\psi_t\> ~|~  t\in \R\}$  coincides with the full set of multiphase covariant states  $\{  |\psi_{\bs \theta}\>  ~|~   \bs \theta \in   [0,2\pi  )^{\times  {d-1}}\}$.  
On the other hand, an $N$-to-$M$ cloner for the clock states will achieve fidelity  
$$F_{\rm prob}  [N\to M]\ge 1-2d\exp\left(-\frac{2p_{\min}^2 N^2}{M}+\frac{4N}{Md}\right)   \, ,  \qquad p_{\min}  \equiv   \min_{j}\,  p_j  $$
 independently of $t$. By taking limits, we then obtain that every multiphase covariant state can be cloned with the same fidelity.   %Hence, the family of multiphase covariant states can be superreplicated. 

Another multiparameter family of states that can be  superreplicated is the family of all maximally entangled states. The  superreplication of maximally entangled states will be discussed in detail in Section \ref{sec:gate}.

It is important to  note that not all quantum states can be superreplicated. First of all, the symmetry of the input ensemble can sometime inhibit superreplication. For instance, if one tries to clone an arbitrary unknown finite dimensional state,  the probabilistic cloner will yield the same fidelity as the optimal deterministic cloner \cite{chiribella-yang-2013-natcomm}.   The lack of probabilistic advantages for universal cloning is an appealing feature from the point of view of the many-worlds interpretation of quantum mechanics \cite{everett-1957-rmp}, and can be even viewed as 
an  axiom: 
 \begin{ax}[Many-Worlds Fairness]\label{prop:many}
 The maximum rate at which a completely unknown state can be cloned is the same in all possible worlds.
\end{ax}
Interestingly, Many-Worlds Fairness rules out the variant of quantum mechanics on real Hilbert spaces, originally considered by Stueckelberg \cite{stueckelberg-1960-hpa} and recently analyzed by Hardy and Wootters \cite{hardy-wootters-2012-foundphys,wootters-2013-arxiv} . The argument to exclude quantum mechanics on real Hilbert spaces is the following:  if the wavefunction had only real amplitudes, then one branch of the wavefunction would allow one  to superreplicate every state of a quantum bit, while the other branches would not,  in violation of the Many-Worlds Fairness. 
The fact that ordinary quantum mechanics allows for complex amplitudes results in the Many-World Fairness property  and in the impossibility of superreplicating completely unknown pure states.  
  
Another limitation to superreplication can be observed for  coherent states of the harmonic oscillator, parametrized as 
\[  |z\>  =  e^{- |z|^2/2} \, \sum_{n=0}^\infty \frac {z^n}{n!} \,  |n\>  \, , \qquad z \in  \C \, . \]
  When $z$ is chosen at random from a Gaussian distribution $p(  z)   =  \lambda/ \pi  \,   e^{-\lambda \, |z|^2} $,  $\lambda >0$, the optimal probabilistic fidelity is given by  \cite{chiribella-xie-2013-prl}
  \begin{align}\label{xie}F_{\rm prob} [N \to M]   =    
  \left\{     
    \begin{array}{ll}
   \frac{  (1+ \lambda)   N}M  \qquad &    \lambda  < \frac M N  -1   \\
   1   &   \lambda    \ge \frac  MN   -1  \, .
  \end{array}
  \right.  
   \end{align}
From the above expression we see that the fidelity vanishes for all replication rates $\alpha > 1$.  This fact is in stark contrast for situation for coherent states with \emph{known} amplitude   \cite{fiurasek-2004-pra,ralph-lund-2009-QCMC}, for which  every rate is achievable.    

Eq. (\ref{xie}) implies that superreplication with rate $\alpha  = 1$ is still possible, provided that $\lambda$ is sufficiently large.  For example, for $\lambda \ge 1$ one can probabilistically transform $N$ copies into $M  = 2 N$ copies, thus  violating the asymptotic no-cloning theorem.   Here it is important to stress  the role of the prior information:  in the lack of it ($\lambda = 0$), the fidelity of the  optimal probabilistic cloner is equal to the fidelity of the optimal deterministic cloner, originally proposed  by Braunstein \emph{et al}  \cite{braunstein-cerf-2001-prl}.   
  
\medskip

\section{The relation between state superreplication and state estimation}\label{sec:compare}

\noindent{\bf Superreplication  via  state estimation.}  A fundamental fact about quantum cloning is its asymptotic equivalence with state estimation \cite{gisin-massar-1997-prl, bruss-ekert-macchiavello-1998-prl, bae-acin-2006-prl,chiribella-dariano-2006-prl, chiribella-2010-tqc, chiribella-yang-2014-njp}:    when the number of output copies becomes  large,   the optimal  cloning performance can be achieved by a classical cloning strategy, which  consists in measuring the input copies and using the outcome of the measurement as an instruction to  prepare the output copies.  In the standard setting, one fixes the number of input copies $N$ and lets the number of ouput copies $M$ go to infinity. For probabilistic cloning, this scenario has been analyzed by Gendra \emph{et al}   \cite{gendra-calsamiglia-2014-prl}, who proved the asymptotic equivalence between probabilistic cloning and probabilistic state estimation.  In the case of clock states that are dense  in an $f$-dimensional manifold, Gendra \emph{et al} showed that the  probabilistic fidelity decays as $F_{\rm prob} [N\to M]  \propto(N^2/M)^{f/2}$ for $M\gg N^2$, while the  deterministic fidelity decays as $F_{\rm det}  [N\to M]  \propto (N/M)^{f/2}$.    An open question is  whether probabilistic estimation also allows to achieve superreplication. 
We now show that this is indeed the case. 

Consider a classical cloning strategy  for the states $  \{  |\psi_x\>~|~x\in\set X\}$, based on probabilistic  estimation of the  parameter $x$ and on the preparation of the state $|\psi_{\hat x}\>^{\otimes M}$, where $\hat x$ is the estimate. 
 This is not the most general classical strategy, but will be sufficient to establish the possibility of superreplication.      The successful quantum operation is given by  
\begin{align}
\map M_{\rm yes}  (\rho)    =     \sum_{\hat x\in\set X}     \,     |\psi_{\hat x}\>\<\psi_{\hat x}|^{\otimes M}  \,    \Tr  [      P_{\hat x}\,   \rho   ] \, ,
\end{align} 
where $  \{P_x\}_{x\in \set    X}$ are positive operators satisfying the normalization condition  
\[  \sum_{x\in\set X}    P_x    +     P_?    =  I  \, ,  \]
for some operator $P_?\ge 0$ associated to  the unsuccessful outcome of the estimation strategy.  
 The fidelity of the strategy can be computed using Eq. (\ref{Fbayes}), which gives 
 \begin{align*}
 F_{\rm prob}  [N\to M] & =   \sum_{x, \hat x}   \,   p(x  |  {\rm  yes})      \,     p( \hat x|  x)   \,  |\<  \psi_{\hat x}|  \psi_x\>  |^{2M} 
 \\
 &  =  \mathbb E  (    |\<  \psi_x|  \psi_{\hat x} \>|^{2M} )  \, , 
 \end{align*}
 $\mathbb E$ denoting the expectation with respect to the probability distribution  $  p(x,\hat x|{\rm yes})  :  = p(x  |  {\rm  yes})      \,     p( \hat x|  x)  $.    Using the convexity of the function $f(y)  =  y^M$, we then obtain the inequality  
 \[  \mathbb E  (    |\<  \psi_x|  \psi_{\hat x} \>|^{2M} )     \ge \left[  \mathbb E  (    |\<  \psi_x|  \psi_{\hat x} \>|^{2} ) \right]^M  \, , \]  
 or equivalently
 \begin{align*}
 F_{\rm prob} [N\to M] &  \ge  \left(  F_{\rm prob}  [ N\to 1] \right)^M  \, , 
 \end{align*}
 meaning that the $M$-copy fidelity of our measure-and-prepare strategy is lower bounded by the $M$-th power of the single-copy fidelity.   Now, suppose that the probabilistic single-copy fidelity approaches 1 as  $1/N^\beta$ for some $\beta  >  0$, namely 
 \begin{align*}
 F_{\rm prob} [N\to 1]    \ge 1   -     O \left(  \frac 1 {N^\beta}\right)  \, ,
 \end{align*}
Then, Bernoulli's inequality yields the bound  
 \begin{align}\label{bernoulli} 
 F_{\rm prob}  [N\to M]  \ge 1   -    O  \left( \frac{ M}{N^{\beta}}  \right)  \, .
 \end{align}
 From the bound it is clear that every replication rate $\alpha  <  \beta$ can be achieved: the fidelity converges to 1 as long as $M(N)/N^\beta$ vanishes in the large $N$ limit.   In summary, we have proven the following
 \begin{theo}\label{theo:beta}
Let $\set S = \{  U_x~|~  x\in\set X\}$ be a continuous set of gates. A probabilistic strategy that estimates the gates  in  $\set S$  with fidelity larger than $1-O(1/N^\beta)$ can be used to replicate the states in $\set S$ at every rate $\alpha  <  \beta$ via a measure-and-prepare strategy
\end{theo}

Theorem \ref{theo:beta}  applies to arbitrary sets of states and to probabilistic strategies with arbitrary constraints on the probability of success.    In the case of clock states, assuming no constraint on the probability of success,  the single-copy fidelity of  phase estimation has the Heisenberg scaling $F_{\rm prob}   [N\to 1]  =  1  -    O\left(1/N^2\right)$  \cite{gendra-ronco-2013-prl}.  Hence, the classical strategy described above achieves superreplication for every replication rate $\alpha  <  2$.    
Note, however, that the classical superreplication strategy has an error vanishing with the power law $1/N^{2-\alpha}$, while the quantum superreplication strategy has an error vanishing faster than every polynomial. 

\medskip 

\noindent {\bf Deriving precision  limits from superreplication.}   We have seen that the replication rate for clock states is determined by the Heisenberg limit in the probabilistic case and by  the standard quantum limit in the deterministic case.    On the other hand, we also seen  strong converse results  (Theorems \ref{theo:strong1} and \ref{theo:strong2}) that prove the optimality of the replication rates without invoking the precision limits of quantum metrology.    In fact,  the quantum metrology limits can be \emph{derived} from our our bounds on the replication rates. The argument is by contradiction:  
\begin{enumerate}
\item Suppose that one could violate the standard quantum limit, using   $N$ copies of the clock state $|\psi_t\>$ to  estimate the parameter $t$ deterministically with error scaling as $1/N^{1+\epsilon}$,  $\epsilon  >  0$. 
 Then, Eq. (\ref{bernoulli}) would imply that one can produce $M$ clones with $M  =   \Theta  (N^{1+\epsilon/2})$, thus violating  Theorem \ref{theo:strong1}. 
 \item Suppose that one could violate the Heisenberg limit of probabilistic metrology, using  $N$ copies  of the clock state $|\psi_t\>$ to  estimate the parameter $t$ probabilistically with error scaling as $1/N^{2+\epsilon}$, $\epsilon  >  0$.   Then, Eq. (\ref{bernoulli}) would imply that one can produce $M$ clones with $M  =   \Theta  (N^{2+\epsilon/2})$, thus violating  Theorem \ref{theo:strong2}.
 \end{enumerate}     
In conclusion, we have shown a complete equivalence between the   limits on the scaling of the error in  quantum metrology and the limits on the replication rate   set by Theorems \ref{theo:strong1} and \ref{theo:strong2}.

\section{Superreplication with reduced interaction size}\label{sec:network}
\noindent{\bf The divide-and-clone approach.} 
To realise the optimal $N$-to-$M$ probabilistic cloner, a global interaction involving at least $M$ systems is required. This can be seen in the example of the equatorial qubit cloner, which is defined as an evolution acting coherently on the basis of the Dicke states, cf. Eq. (\ref{cloner}).    Of course, the need for interactions among large numbers of systems makes the  implementation of cloning a challenging issue.   Here we investigate the possibility to reduce the scale of the interactions.   

A simple strategy to reduce the interaction scale is a ``divide-and-clone" strategy, where one divides the input copies into groups and performs optimal cloning  on each group. Suppose that  we divide the $N$ input copies into groups of $N':=  \Theta (N^\beta)$ copies. Applying the optimal $N'$-to-$M'$ probabilistic cloner to each individual group, we then achieve fidelity 
$$F  [N'\to M']\ge 1-2K\exp\left[-\frac{2p_{\min}^2 (N')^2}{M'}+\frac{4N'}{M'K}\right]$$ for each group. The value of $M'$  can be chosen in order to reach the desired replication rate:  in order to obtain $M$ copies overall, one needs $M'$ to satisfy the condition $M  =  M'  \,  N/ N'$.  For a replication process producing $M  =   \Theta( N^{1+\delta})$ copies, the condition yields    $ M'    =    \Theta  (N^{\delta+\beta})$.  

Since there are $N/N' $ groups, the overall fidelity of this strategy is  
\begin{align*}
F  [N\to M] & = (F[  N'\to M'])^{N/N'}\\  
 &  \ge    1-    \Theta( N^{1-\beta})  \,  \exp\left[-  \Theta  \left( N^{\beta  -\delta}  \right)    \right].\end{align*}
From the above bound it is immediate to see that the fidelity converges to 1 whenever $\beta>\delta$. 
Hence, superreplication of quantum clocks can be achieved with interactions among $M'=O\left(N^{2\delta+  \epsilon}\right)  =    O  \left(   M^2/N^{2-\epsilon} \right)$ particles, for every desired $\epsilon  >0$.  Now, recall that  the original superreplication strategy \cite{chiribella-yang-2013-natcomm} requires interactions among $O(M)$ particles.  This means that  the modified strategy leads to a reduction of the interaction  scale by a factor  $N^{2-\epsilon}/M$.  

On the other hand, the modified strategy does not allow to achieve superreplication with interactions among  less than $\Theta(  N^\delta )$ systems.   Indeed,  superreplication can only be achieved if  the fidelity of the individual processes approaches 1. By the Heisenberg limit, this condition is satisfied only if   $N'$ and $M'$ satisfy the asymptotic  relation  $M'\ll    (N')^2$.      Recalling that $M'$ has to scale as $  N^{\beta+ \delta}$ and that $N'$ scales as $N^\beta$, we then obtain that $\beta  >  \delta$ is a necessary condition.  In summary,  the strategy of dividing the input copies into non-interacting groups achieves superreplication with $M  =    \Theta  (N^{1+\delta})$ output copies if and only if the size of each group grows faster than $N^\delta   =  M/N$.

\medskip
\noindent{\bf A sequential cloning approach.}   Another approach to reduce the scale of interactions is to generate clones by local mechanisms.    For example, we can imagine a cloning process that involves repeated interactions among $O(N)$ systems.  Let us analyze this idea in the case of equatorial qubit states: here we 
consider a sequence of cloners that take $N$ to $2N$ copies, arranged in the following circuit:
\begin{equation*}
\begin{aligned}\Qcircuit @C=1em @R=0.2em @!R
{
   & \multigate{1}{   {\map{M}^{(N)}}} & \qw&\qw& \qw&\qw  &\qw&\qw&\qw &\qw\poloFantasmaCn{1}&\qw\\
  & \pureghost{  {\map{M}^{(N)}}  } &\qw&\qw &\multigate{1}{  {\map{M}^{(N)}}  } &\qw&\qw&\qw&\qw &\qw\poloFantasmaCn{2}&\qw \\
  & & & &\pureghost{  {\map{M}^{(N)}}} & \qw&\qw&\\
  & & & & && \vdots&\\
  & & & & & & &\multigate{1}{  {\map{M}^{(N)}}  }&\qw&\qw\poloFantasmaCn{K-1}& \qw  \\
  & & & & & & &\pureghost{ {\map{M}^{(N)}}} &\qw& \qw\poloFantasmaCn{K} &\qw
}
\end{aligned}
\end{equation*}
Each wire in the circuit represents a composite system of $N$ qubits.   At each step,  the quantum operation $\map M^{(N)} $  performs the optimal  $N$-to-$2N$ probabilistic cloning,  given by the map 
\begin{align*}
  \left |N,  \frac N2+m  \right \> \quad  \longrightarrow \quad  \sqrt{\frac{  {{2N\choose N+m}} }{{{2N\choose\frac{3N}2}{N\choose\frac N2+m}}  }}   \,  \left |2N,N+m  \right\>   \, , \qquad  \,  \forall m\in\left [-\frac N2,  \frac N2 \right] \, ,  
\end{align*}
as one can see substituting $M$ with $2N$ in  Eq. (\ref{cloner}). In the following it will be convenient to  express the map as 
\begin{align*}
  \left |N,  n  \right \> \quad  \longrightarrow \quad  \sqrt{\frac{  {{2N\choose N/2+n}} }{{{2N\choose\frac{3N}2}{N\choose n }}  }}   \,  \left |2N, \frac N 2+n  \right\>   \, , \qquad  \,  \forall n\in\left [  0  ,  N \right] \, ,  
\end{align*}

We now analyze the performance of the sequential cloner after $K-1$ steps, which result into the generation of $KN$ approximate copies. 
At the first step, the input state is the $N$-copy state \begin{align*}
|\Psi_{t,0}\>=\frac{1}{2^{N/2}}\sum_{n=0}^{N}\sqrt{{N\choose n}}e^{-in t}  \,   \left|N,     n  \right\>_{1} \, ,
\end{align*}
where the subscript $1$  indicates the first group of $N$ qubits. 
\begin{figure}[t!]
\centering
\includegraphics[width=0.55\linewidth]{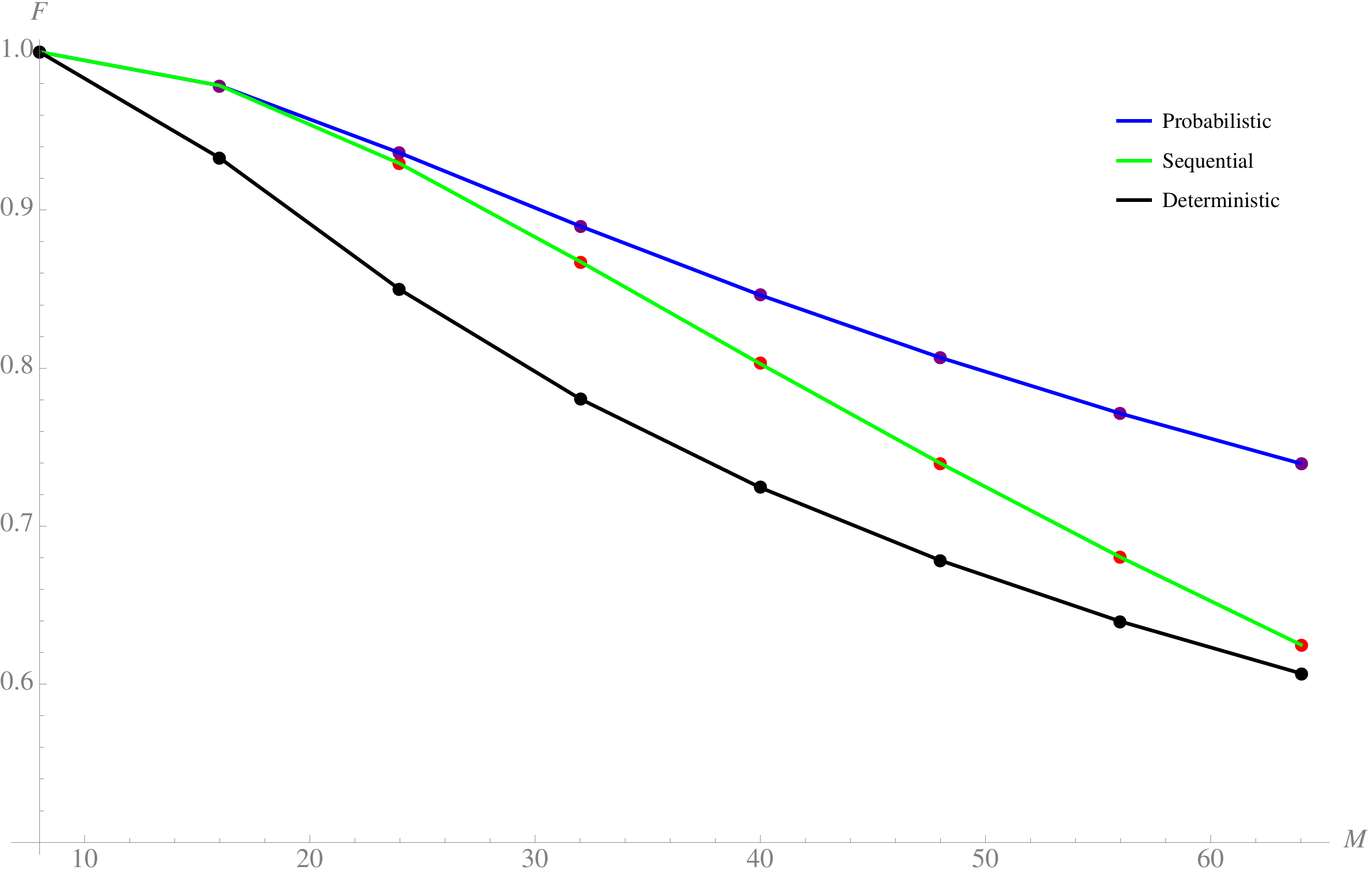}\\
\caption{\noindent{\bf Comparison between the three cloners.} In this figure we compare the performance of the network cloner to the one of the probabilistic cloner as well as the one of the deterministic cloner. The fidelity is plotted as a function of the output number, fixing the input number to $N=8$.  The green line (with numerics represented by red dots) represents the fidelity-output curve for the network cloner.  The blue line (with numerics represented by red dots) stands for the fidelity-output curve for the probabilistic cloner, and the black line (with numerics represented by black dots) stands for the fidelity-output curve for the deterministic cloner.}
\label{fig:comparison}
\end{figure}
The first  cloner transforms $|\Psi_{t,0}\>$ into a state on the first and second group, yielding  $2N$ qubits  in the state
\begin{align*}
|\Psi_{t,1}\>&=N_1\sum_{n=0}^{N}\sqrt{2N\choose N/2+n} \,  e^{-in t }  \,  \left |2N, \frac N 2 +n\right\>_{1,2}\\
&=N_1\sum_{n=0}^{N}e^{-i n t }\sum_{x_1\in\set{S}_{n,1}}\sqrt{{N\choose n-x_1}{N\choose N/2+x_1}} \, \left |N, n-x_1\left\>_{1}  \left |N,\frac N2 +x_1\right\>_{2}  \right. \right. \, ,
\end{align*}
where $N_1$ is a normalization constant and the set $\set S_{n,1}$ is defined via the relation 
\begin{align}
\set{S}_{n,K}=\left\{  \st x  :  =  (x_1,\dots,x_k)~\left |~x_j\in  \left[-\frac N2,  \frac N2 \right],  \quad  \sum_{l=1}^{j}  \, x_ l\in[n-N,n] \, ,   \quad \forall\, j\in   \{1, \dots , K \}  \quad   \right\}  \right. \,.
\end{align}
Iterating the cloning process for $K-1$ steps we finally obtain $K$ groups of qubits in the joint state 
\begin{align*}
\left|\Psi_{t,K-1}\right\>=N_{K-1} \sum_{n=0}^{N}e^{-int }\sum_{\set x \in\set{S}_{n,K-1}}  \,  
  \,    c_{n,\st x}  \,   \left|N,n-\sum_{i=1}^{K-1}  \,  x_i\right\>_{1}      \left |  N,  \frac N2  +  x_1  \right\>_2 \cdots       \left |  N,  \frac N2  +  x_{K-1}  \right\>_K \, ,
 \end{align*}
where  $c_{n, \st x}$ is the coefficient given by  
\[  c_{n, \st x}  :  =  \sqrt{   {N\choose {n-\sum_{i=1}^{K-1}x_i    }}  \,    \prod_{j=1}^{K-1}     {N\choose {N/2+x_j}}}           \]

The overall fidelity of the cloner with the state of $NK$ perfect copies does not depend on $t$ and is given by 
\begin{align}\label{Fnet}
F_{\rm seq}   [  N\to KN] =\frac{1}{2^{NK}}\sum_{n=0}^{N}  \,  \sum_{\st x \in\set{S}_{n,K-1}}\left[\prod_{j=1}^{K-1}{N\choose N/2+x_j}\right]{N\choose n-\sum_{i=1}^{K-1}x_i}.
\end{align}
For $K>2$, the cloning fidelity (\ref{Fnet}) is strictly smaller than the  fidelity (\ref{Fprob}) of the optimal probabilistic cloner. 
A comparison of the sequential cloner, the optimal probabilistic cloner and the optimal deterministic cloner is illustrated in Fig. \ref{fig:comparison} for $N=8$.     Note that the fidelity of the sequential cloner  decays quickly with the number of cloning steps. In our example, the fidelity of the sequential cloner approaches the fidelity of the deterministic cloner as the number of output copies approaches $M  = N^2 = 64$.   This numerical result suggests that it may not be possible to achieve superreplication sequentially. However, it is an open question whether or not superreplication can be achieved by the the sequential cloner or by similar mechanisms.    For example, one could construct a quantum cellular automaton for cloning, as proposed  by D'Ariano, Macchiavello and Rossi in \cite{dariano-macchiavello-2013-pra}. They considered a tree-shaped network of   deterministic  1-to-2  cloners.   In order to achieve super-replication, one would have to extend the model, allowing for probabilistic cloners. Moreover,  symmetry arguments imply that probabilistic 1-to-2 cloners do not offer any advantage over their deterministic counterparts \cite{chiribella-yang-2013-natcomm}.  Hence, one has to consider local cloners with larger number of input copies.  An interesting possibility is to construct a probabilistic cellular automaton where the building blocks are the optimal $k$-to-$2k$ probabilisitc cloners with $k>1$.

\section{Quantum gate superreplication}\label{sec:gate}

%\subsection{Gate superreplication network}

We have seen that state superreplication has necessarily a vanishing success probability in the asymptotic limit.  Here we analyze the task of replicating quantum gates, where  superreplication can be achieved with unit probability of success on average over all input states.

\begin{figure}[t!]
\centering
\includegraphics[width=0.6\linewidth]{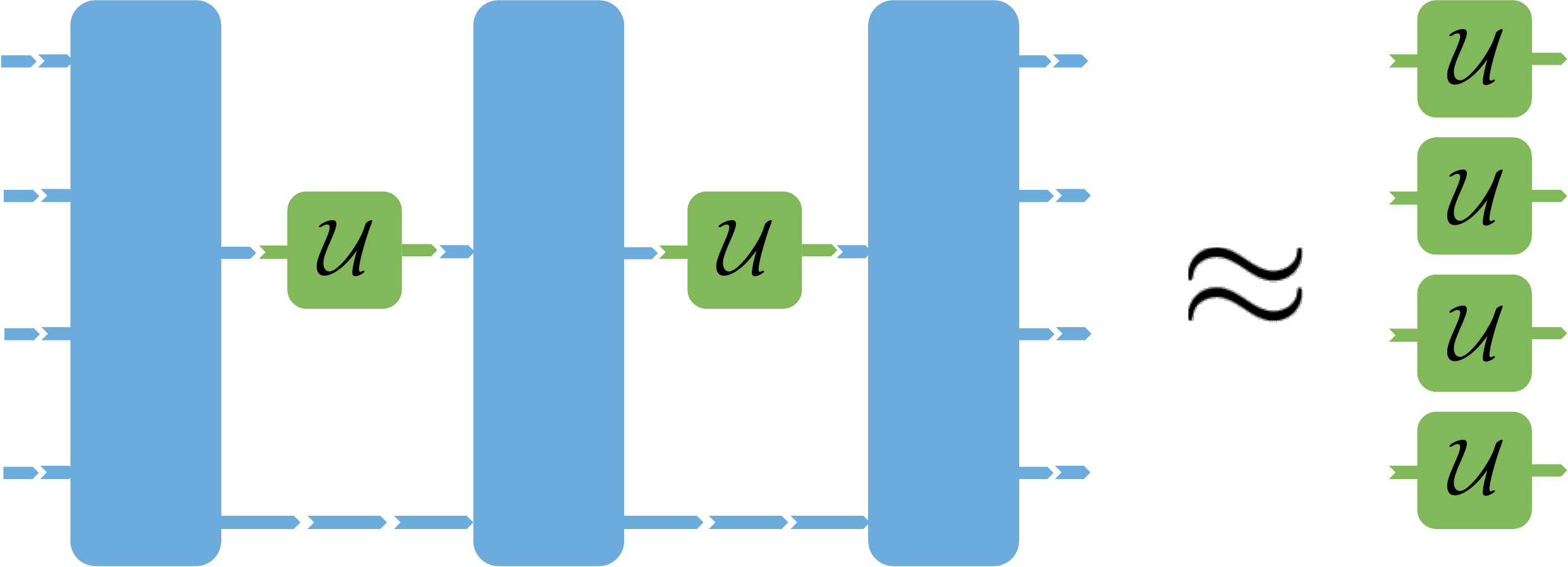}\\
\caption{\noindent{\bf Quantum gate cloning.} A quantum network for gate cloning. Given $N$ uses of an unknown unitary $\map{U}$ (green boxes on  the left), the network (blue boxes on the left)  simulates $M>N$ uses of $\map{U}$ (green boxes on the right).  }
\label{fig:network}
\end{figure}

\medskip
\noindent{\bf Gate superreplication.}    In many applications, including quantum metrology and quantum algorithms, one is given access to a black box implementing an unknown quantum gate. In these applications the uses of the gate are a resource: indeed, a  no-cloning theorem for gates  asserts  that it is impossible to perfectly simulate two uses of an unknown gate  by using it only once \cite{chiribella-dariano-2008-prl}. 
% In other words, it is impossible to duplicate the uses of an unknown quantum gate. 
 Still, a natural question is: how well can we simulate $M$ uses of the unknown gate with $
N<M$ uses?      Here  the problem is to engineer a  quantum computational network that uses the unknown gate as a subroutine, as in figure \ref{fig:network}.   In analogy with the superreplication of quantum states, we say that  superreplication of quantum gates is possible iff one can find a sequence of  networks with the property that the simulation error vanishes in the asymptotic limit and the number of extra copies of the input gate grows as $N$ or faster.  In the following we will see that gate superreplication can be achieved deterministically for almost all input states. 
The first result of this type was discovered by D\"ur, Sekatski, and Skotiniotis \cite{dur-sekatski-2015-prl} in the case of phase gates, that is,  gates of the form \[U_t =\sum_{n=0}^{d-1}e^{-i  \,  n  t } \, |n\>\<n| \, , \qquad  \forall t \in[0,2\pi) \, . \] 
The performance of the simulation was quantified by  the  fidelity between the output state of the simulation and the output state of $M$ perfect uses of the gate  $U_t$, averaged over $t $ and over all possible input states. 
%\begin{align}\label{F-gate}
%F_{\rm gate}=\int \frac{\d \theta}{2\pi} \, \int \d\Psi ~ \<\Psi|(U_\theta^\dag)^{\otimes M}\map{C}_\theta \left(|\Psi\>\<\Psi|\right)U_\theta^{\otimes M}|\Psi\> \, , \qquad |\Psi\>\in\spc{H}^{\otimes M} \, , 
%\end{align}
%
Using a deterministic network, the authors showed how to simulate $M\ll N^2$ uses with asymptotically unit fidelity.  
This result can be extended from phase gates to arbitrary gates  \cite{chiribella-yang-2015-prl}, as discussed in the following paragraphs.

\medskip
\noindent{\bf Universal gate superreplication.}    Given $N$ uses of a unitary gate, the goal is to  simulate $M$ parallel uses of the gate.   Let us denote by $\map{C}^{(N)}_U$ the quantum channel that results by inserting $N$ uses of the unknown gate in the cloning network.  With this notation, the average  fidelity reads 
\begin{align}\label{F-gate}
F_{\rm gate}  [N\to M]=\int \d U \, \int \d\Psi ~ \<\Psi|(U^\dag)^{\otimes M}\map{C}^{(N)}_U\left(|\Psi\>\<\Psi|\right)U^{\otimes M}|\Psi\> \, , 
\end{align}
where $\d U$ is the normalized Haar measure over the group of all unitary gates $\d \Psi$ is the normalized Haar measure over the manifold of pure $M$-partite states.      By Markov' inequality,   
  a simulation with gate fidelity  $F_{\rm gate}  [N\to M]  \ge 1-\epsilon$ is a simulation that works with fidelity $F\ge 1-\delta$ on all pure states except for a small fraction, whose probability is  smaller than $\epsilon/\delta$. 
  
A replication process is described as a sequence of networks, with the $N$-th network transforming $N$ copies of the unknown gate into $M(N)$ approximate copies. We say that the replication process is \emph{reliable} iff   
\[\lim_{N\to\infty} F_{\rm gate} [N\to M(N)]=1 \, .\] 
In other words, for a reliable replication process one has $F[  N  \to M(N)]  \ge 1-  \epsilon_N$ for some $\epsilon_N$ converging to zero.  Choosing $\delta_N   = \sqrt{ \epsilon_N}$, we then have that the replication process simulates the desired gate with fidelity larger than $1-\epsilon_N$ on all  pure states except a small  fraction, whose probability is smaller than  $\epsilon_N/\delta_N    =  \sqrt{  \epsilon_N}$.   As in the case of state replication, we say that a process has \emph{replication rate} 
 \begin{align*}
 \alpha  =  \liminf_{N\to \infty} \frac{ \log   [M(N)  -  N]}{  \log N}  \, 
 \end{align*} and we say that the rate $\alpha$ is \emph{achievable} iff  there exists a reliable replication process with that rate. 
 
 In the following, we will see that every rate smaller than $2$ is achievable.  
For simplicity, we discuss the case of qubit gates, although the protocol can be extended in a straightforward way to  unitary gates on aribtrary finite-dimensional quantum systems.

 The key to  the construction of the universal gate cloning network is the decomposition of  the Hilbert space of $K$ qubits into orthogonal subspaces corresponding to different values of the total angular momentum. In formula, one has 
 \begin{align*}
\spc{H}^{\otimes K}\simeq\bigoplus_{j=0}^{K/2}\left(\spc{R}_j\otimes \spc{M}_{jK}\right) \, ,
\end{align*}
where $j$ is the quantum number of the total angular momentum, $\spc R_j$ is a representation space  and $\spc M_j$ is a multiplicity space \cite{book-fulton-harris-representation}. 
%The isomorphism in the above equation is called the Schur transform and has a polynomial-time  implementation \cite{schur-transform}.
With respect to this decomposition, the action of  $K$ parallel uses of a generic qubit unitary $U$ has the block-diagonal form
\begin{align*}
U^{\otimes K}\simeq\bigoplus_{j=0}^{K/2}\left(U_j\otimes I_{jK}\right) \, ,
\end{align*}
where $  I_{jK}$ is the identity on the multiplicity space $\spc M_{jK}$. 

Note that $U^{\otimes K}$ acts non-trivially only in the representation spaces. This means that, essentially, the multiplicity spaces can be eliminated without losing any information about $U$. Likewise, they can be inflated by adding  ancillas. 
  The working principle for the simulation of $U^{\otimes M}$ given $U^{\otimes N}$ will be to conveniently adjust the size of the multiplicity spaces, as in the following
  \begin{protocol}[Gate superreplication network \cite{chiribella-yang-2015-prl}]\label{protocol:gate} 
To construct the replication network, one has  to %The replication network is constructed following the procedure below: 
\begin{enumerate}
\item Choose an ancilla $A$ such that  the rank of $I_{jN}\otimes I_A$ is no smaller than the rank of $I_{jM}$ for every $j\in[0,N/2]$;
\item Compose the gate  $U^{\otimes N}$  with  the identity   on $A$, thus obtaining the gate  $U^{\otimes N }  \otimes I_A  =  \bigoplus_{j=0}^{N/2}(U_j\otimes I_{jN}\otimes I_{A})$; 
\item Project the resulting gate $U^{\otimes N}\otimes I_A$ into the subspace $$\spc{H}_{N}=\bigoplus_{j=0}^{N/2}\left(\spc{R}_j\otimes \spc{M}_{jM}\right)  \subset  \spc H^{\otimes N}  \otimes \spc H_A$$ where the gate acts as 
\begin{align*}
U'  =  \bigoplus_{j=0}^{N/2}\left(U_j\otimes I_{jM}\right) \, .
\end{align*}
\item Embed the subspace $\spc H_N$ into the Hilbert space $\spc H^{\otimes M}$. With a slight abuse of notation we  use $\spc H_N$ both for the subspace of $\spc H^{\otimes N}  \otimes \spc H_A$ and for the corresponding subspace of $\spc H^{\otimes M}$. 
\item Given an input state of $M$ qubits,  perform a projective measurement that projects the qubits either in the subspace $\spc H_N$ or in its orthogonal complement $\spc H_N^\perp$.  If the qubits are projected into $\spc H_N$, then apply the gate $U'$.  If the qubits are projected into  $\spc H_N^\perp$, then perform the identity.   
\end{enumerate}
\end{protocol}
The  network described by protocol \ref{protocol:gate} perfectly imitates $U^{\otimes M}$ for any state in the   subspace $\spc{H}_{N}  \subset \spc H^{\otimes M}$.  Now,  it is possible to show that a random pure state in $\spc{H}^{\otimes M}$ has a large overlap with the subspace $\spc H_N$ whenever  $M\ll N^2$.   Indeed, we have the following 
\begin{lemma}\label{lemma:concentration}
Let $F_N$ be the fidelity between the pure state $|\Psi\>\in\spc{H}^{\otimes M}$ and its projection $|\Psi_N\>   =   P_n |\Psi\>/\|  P_N  |\Psi\>\|$.  The expectation value of the fidelity over the uniform measure is given by  
\begin{align*}
\mathbb E  \left(  F_N\right)   &  =  \int \d \Psi    \,     \<\Psi|   P_N   |\Psi\>     \\
&  =   \frac{ \Tr  [  P_N]}{2^M}  \\
& =  \sum_{j=0}^{N/2}\frac{d_j m_{jM}}{2^M}\\
&    \ge  1  -   (M+1)  \, \exp\left[-\frac{N^2}{2M}\right]  \, ,
\end{align*}  
\iffalse 
The probability that $F_N$  is smaller than $1-\epsilon$ satisfies the bound
$$\mathbf{Prob}[F_{N}<1-\epsilon]<\frac{(M+1)}{\epsilon}\exp\left[-\frac{N^2}{2M}\right]$$
for any $\epsilon>0$.
\fi
\end{lemma}
In the above inequality $d_j$ and $m_{jM}$ are the dimensions of the representation space $\spc{R}_j$ and the multiplicity space $\spc{M}_{jM}$, respectively. The detailed proof of the last inequality can be found in the supplemental material of Ref. \cite{chiribella-yang-2015-prl}. Using the above lemma, it is easy to show that the fidelity is close to 1 whenever $M $ is small compared to $N^2$.  Indeed,
 the cloning fidelity   (\ref{F-gate}) is lower bounded as 
\begin{align*}
F_{\rm gate}  [N\to M] & \ge  \int \d U \, \int  \d\Psi ~ \<\Psi|(U^\dag)^{\otimes M}  \,   U'    P_N  |\Psi\>\<\Psi|  P_N  \,  U^{'\dag}   U^{\otimes M} \, |\Psi\>    \\
& =  \int  \d\Psi ~   (\<\Psi|     P_N  |\Psi\>)^2 \\
&  \ge  \left(  \int  \d\Psi ~   \<\Psi|     P_N  |\Psi\>\right)^2    \\
&  =   \left[   \mathbb E   (F_N)  \right]^2  \\
&  \ge  1 -  2      (M+1)  \, \exp\left[-\frac{N^2}{2M}\right]    \, .  
\end{align*} 
In conclusion, the gate fidelity converges to 1 for all replication processes with replication rate $\alpha <2$, establishing the possibility of gate superreplication.       Similar bounds can be found for  arbitrary $d$-dimensional systems, in which case we have  
\begin{align}\label{bound:dimensiond}
   F_{\rm gate }  [N\to M]  \ge 1-2(M+1)^{\frac{d(d-1)}{2}}\exp\left[-\frac{N^2}{2M}\right]\qquad\forall d<\infty.
\end{align}
The inequality follows from the concentration of the Schur-Weyl measure \cite{alicki-rudnicki-1988-jmp,meliot-2010-arxiv}, which has applications in quantum estimation \cite{marvian-spekkens-2014-cmp} and information compression \cite{thesis-harrow-2006,yang-ebler-2015-arxiv}.   
 The bound (\ref{bound:dimensiond}) allows one to produce $M  = \Theta (N^{2-\epsilon}) $  copies with a super-polynomially fast vanishing error. In addition, one can produce $M  =   \Theta [N^2/\log N]$  copies with an error vanishing faster than $1/N^k$ for every desired $k>0$. For this purpose, it is enough to choose 
 \[M =  \left \lfloor \frac{ N^2}{2 (d^2-d+k)  \,  \log N  }\right\rfloor \, ,\] 
 so that Eq. (\ref{bound:dimensiond}) becomes
 \begin{align}\label{forestimation} F_{\rm gate }  [N\to M]  \ge 1- O\left(    \frac1 {  N^k  \,   \log N^{d(d-1)/2}} \right)  \, , \qquad   N  \gg d \, . 
 \end{align}

As a final remark, we note that the size of the interactions used in the superreplication network can be reduced from  $O(M)$ qubits to $O (  M/N )$ qubits, by dividing the $N$ input gates into groups of $O( N^2/M)$ gates.     This fact can be proven with the same argument used for quantum states.   

\section{Applications of gate superreplication}

\noindent{\bf Superreplication of maximally entangled states.}\label{sec:application}
Inspired by gate superreplication,  one can construct a protocol for superreplicating maximally entangled states, which achieves any cloning rate $\alpha<2$ \cite{chiribella-yang-2015-prl}.   For qubits, we can cast an arbitrary maximally entangled state in the form $|\Phi_U\>  =(U\otimes I)|\Phi^+\>$ where $|\Phi^+\>$ is the Bell state $|\Phi^+\>=(|00\>+|11\>)/\sqrt{2}$.    The replication protocol works as follows: First,  the unitary gate $U$ can be extracted from the state via a gate teleportation protocol \cite{gottesman-chuang-1999-nature}. The success probability of a single gate extraction is $1/4$, which implies that $U^{\otimes N}$ can be extracted from $N$ copies of the maximally entangled state with probability $(1/4)^N$.   The extracted gates are then used in the gate superreplication network, which  simulates  the gate $U^{\otimes M}$ with vanishing error whenever $M$ grows as  $N^\alpha$, $\alpha <  2$. Finally, the approximation of $U^{\otimes M}$ is applied locally on  $M$ copies of the Bell state $|\Phi^+\>$, thus providing $M$ approximate copies of the state $|\Phi_U\> $.  The fidelity of the replicas can be bounded as 
\begin{align*}
F_{\rm ent}  [N\to M]&=\int \d U ~ \<\Phi^+|^{\otimes M}(U^\dag\otimes I)^{\otimes M}\left(\map{C}^{(N)}_U\otimes \map{I}\right)\left(|\Phi^+\>\<\Phi^+|^{\otimes M}\right)(U\otimes I)^{\otimes M}|\Phi^+\>^{\otimes M}\\
%&=\int \d U ~ \<\Phi^+|^{\otimes M}\left(\map{U}^{\dag\otimes M}\circ\map{C}^{(N)}_U\otimes \map{I}\right)\left(|\Phi^+\>\<\Phi^+|^{\otimes M}\right)|\Phi^+\>^{\otimes M}\\
&=\frac{(2^M+1)F_{\rm gate}  [N\to M]-1}{2^M} \, ,
\end{align*}
having  used the relationship between the entanglement fidelity and the gate fidelity  derived by Horodecki et al \cite{horodecki-horodecki-1999-pra}. The above equality implies that the fidelity of the state cloning protocol goes to one  if and only if the fidelity of gate superreplication goes to one. Therefore, the above protocol is able to replicate maximally entangled states at every  rate $\alpha<2$. 
%It should be stressed that the protocol   has an exponentially vanishing probability of success, since it involves $N$ independent rounds of  gate teleportation.

The idea of gate teleportation followed by gate superreplication can also be applied to achieve superreplication of other families of states. For instance, gate superreplication yields is an alternative superreplication protocol for equatorial qubit states. Given $N$ copies of an equatorial qubit state $(|0\>+e^{-i  t}|1\>)/\sqrt{2}$, one can first generate $N$ copies of the maximally entangled qubit state $(|00\>+e^{- i t }|11\>)/\sqrt{2}$ by applying a CNOT gate to each input copy. Then,    the maximally entangled qubit state can be superreplicated. Finally, applying a  CNOT gate to each output copy  allows to transform the (approximate) copies of the maximally entangled qubit state into (approximate) copies of the equatorial qubit state.  The net result of this protocol is the superreplication of equatorial qubit states.

Similar arguments apply to the superreplication of maximally entangled states in arbitrary finite dimensions.   Every maximally entangled state can be parametrized as $|\Phi_U\>  =   (U\otimes I )  |\Phi^+\>$,  where $|\Phi^+\>  =   1/\sqrt{d} \,  \sum_{n=0}^{d-1}  \,  |n\>|n\>$.    Again, the unitary gate $U$ can be probabilistically extracted from the state  $|\Phi_U\>$, with probability $p=  1/d^2$.    Hence, superreplication of maximally entangled states can be achieved by %\emph{i)}  extracting $N$ uses of the gate $U$ from $N$ copies of the state $|\Phi_U\>$, \
\emph{i)} simulating $M$ copies of the gate $U$  and \emph{ii)} applying the simulated gates locally on $M$ copies of the state $|\Phi^+\>$.     
The fidelity is then given by 
\begin{align}
\nonumber F_{\rm ent}  [N\to M]&=\int \d U ~ \<\Phi^+|^{\otimes M}(U^\dag\otimes I)^{\otimes M}\left(\map{C}^{(N)}_U\otimes \map{I}\right)\left(|\Phi^+\>\<\Phi^+|^{\otimes M}\right)(U\otimes I)^{\otimes M}|\Phi^+\>^{\otimes M}\\
%&=\int \d U ~ \<\Phi^+|^{\otimes M}\left(\map{U}^{\dag\otimes M}\circ\map{C}^{(N)}_U\otimes \map{I}\right)\left(|\Phi^+\>\<\Phi^+|^{\otimes M}\right)|\Phi^+\>^{\otimes M}\\
&=\frac{(d^M+1)F_{\rm gate} [N\to M]-1}{d^M}  \, , \label{fidd}
\end{align}
again, having  used the relationship between the entanglement fidelity and the gate fidelity   \cite{horodecki-horodecki-1999-pra}.   In conclusion, the cloning fidelity for maximally entangled states  approaches 1 if and only if the  cloning fidelity for unitary gates  approaches 1.   Hence, the superreplication of quantum gates implies the (probabilistic) superreplication of maximally entangled states.  In turn, the supereplication of maximally entangled states implies the superreplication of the uniform-weight multiphase states  \[|\psi_{\bs \theta}\>=    \frac 1{\sqrt {d} } \,\left(    |0\>  +  \sum_{j=1}^{d-1}  \, e^{i\theta_j}|j\> \right)  \, , \qquad \theta_j  \in  [0,2\pi)  \, ,  \forall   j  = 1,\dots,  d-1  \, , \]    
    which are unitarily equivalent to the  maximally entangled states   $|\Phi_{\bs \theta}\>=    \frac 1{\sqrt {d} } \,\left(    |0\>  |0\>+  \sum_{j=1}^{d-1} \, e^{i\theta_j}|j\>|j\> \right)$. 
    
 \medskip  
 
 \noindent {\bf Upper bound on the rate of gate replication.}     
The relation between gate replication and replication of maximally entangled states can be used to derive the ultimate limit on the rates of gate replication in finite dimensions.  The argument proceeds by contradition:  Suppose that it is possible to devise a network  that simulates $\Theta(N^\alpha)$ uses of $U$ for $\alpha\ge2$ with vanishing error. Then we could use it to achieve superreplication of maximally entangled states at rate $\alpha\ge2$.   Now, the family of maximally entangled states contains the family of clock states 
\[  |\Phi_t \>   =      (U_t \otimes I  )   \,  |\Phi_+\>  \, , \qquad   U_t  =  \sum_n \,  e^{-i  n t} \,        |n\> \<n|     \, .   \]   
Replication of these states  at rate $\alpha  \ge 2$  would contradict  Theorem \ref{theo:strong2},   which implies that every  cloner with  rate  $\alpha  \ge 2$ must have vanishing fidelity.  Hence, the possibility of achieving gate superreplication at rates larger than quadratic is excluded.   Note that the above argument applies not only to deterministic gate replication networks, but also to probabilistic networks.   Using the correspondence between entanglement fidelity and gate fidelity,  we obtain  the following:  
\begin{theo}\label{theo:strong4}
Every physical process that replicates phase gates  with replication rate $\alpha  >2$  will necessarily have vanishing gate fidelity (no matter how small  the probability of success). 
\end{theo}
 An alternative  optimality proof for the quadratic replication rate   was presented by Skotiniotis,  Sekatski, and D\"ur \cite{sekatski-skotiniotis-2015-pra} who devised an argument to bound the replication fidelity based on the no-signalling principle.   

\medskip
\noindent{\bf Supergeneration of maximally entangled states.} Gate superreplication can be achieved  deterministically, while state superreplication can only be achieved  with a vanishing probability. The origin of this sharp difference  is in the fact that states and gates are inequivalent resources: while gates can be used to deterministically generate states as  $|\psi_U\>  =  U  |0\>$,    the converse process  is forbidden by the no-programming theorem \cite{nielsen-chuang-1997-prl}.  
  It is then useful to distinguish between the task of state cloning, where the input consists of $N$ copies of the state $|\psi_U\>  =  U  |0\>$, and the task of \emph{state generation}, where the input consists of $N$ copies of the gate $U$.     Deterministic gate superreplication cannot be used to achieve deterministic state superreplication, but   can be used to achieve deterministic state \emph{supergeneration},  that is, the generation of up to $N^2$ almost perfect copies of the quantum state $|\psi_U\>$ from $N$ copies of the gate $U$.      
  A general  supergeneration protocol works as follows:
  \begin{enumerate}
\item   Use a  gate superreplication protocol to simulate $M\ll N^2$ uses of the gate $U$
\item Applying the simulated gates on the  state $ |0\>^{\otimes N}$. 
\end{enumerate}
In general, the protocol can be tailored to the specific set of states that one wants to generate.  For example, one could have \emph{i)} clock states, where $U$ is of the form $U  =  e^{-it H}$, \emph{ii)} maximally entangled states, where
  $U$ is of the product form $U  =   V_A\otimes I_B$ with respect to some bipartition of the Hilbert space, and \emph{iii)} arbitrary pure states, where $U$ is a generic unitary. 
  
The fidelity of supergeneration depends on the protocol used to replicate the gates.  For example, the phase gate replication by D\"ur \emph{et al} allows to supergenerate clock states \cite{dur-sekatski-2015-prl}, while our universal gate replication allows to supergenerate maximally entangled states of bipartite systems \cite{chiribella-yang-2015-prl}.  
Interestingly, the universal gate replication   does not work for the set of all pure states, parametrized as  
\[   \{   |\psi_U\>    =   U  |0\> ~|~       U\in\grp {SU}(d)  \, ,     |0\>  \in  \spc H  \} \, . \]  
In this case, the approach of simulating $M$ uses of the gate $U$ and applying it to the state $ |0\>^{\otimes N}$  does not work, because the state  $ |0\>^{\otimes N}$  lies in the   subspace $\spc{H}_{N}^{\perp}$ where our gate superreplication network fails. However, we will see in a couple of paragraphs that supergeneration  of arbitrary pure states can be achieved  via a suitable protocol based on gate estimation.   

\medskip 

\noindent{\bf Gate estimation with quasi-Heisenberg scaling.}     The supergeneration of maximally entangled states has  an elegant application to quantum metrology. Specifically, it allows for an easy proof of the fact that an unknown quantum gate can be estimated an error scaling as $\log N/N^{2}$.   Here the error is defined as $\<e\>_N  =  1  -   F_{\rm est, gate} [N]$, where $F_{\rm est, gate} [N]$ is the fidelity of gate estimation with $N$ copies, namely
\begin{align}\label{fgate}  
   F_{\rm est , gate}   [N] =   \int \d U  \,  \int \d \hat U   \,   p_N(  \hat U|  U)  \,      F_{\rm gate}   (  \hat U,  U)  \, ,  
   \end{align}
 $p_N(\hat U|U)$ being the probability distribution resulting from the estimate  and $F_{\rm gate} (\hat U, U)$ being the gate fidelity, defined as 
 \[F_{\rm gate}  (\hat U ,  U)   :  =    \int \d \psi  \,    \left|  \<  \psi  |   U^\dag \hat U   |\psi\>  \right|^2 \, . \]
 
 We now show that gate superreplication can be used to achieve  estimation fidelity scaling as 
 \begin{align}\label{gateestimation}
F_{\rm est, gate}   [N]   \ge  1   -    O\left( \frac {\log N}{ N^{2}}  \right)    \, .
\end{align}
  The proof goes as follows:  given $N$ copies of the gate $U$, we can produce $M   =  \Theta(    N^{2}/\log N)$ copies of the maximally entangled state $|\Phi_U\>$, with an error vanishing faster than  $1/N^k$, for every desired $k>0$---cf. Eq. (\ref{forestimation}).    In particular, we set $k=2$.   With this choice,  the error vanishes faster than $1/M$ and the $M$ approximate copies can be used for state estimation, resulting into an estimate of the maximally entangled state with fidelity 
  \begin{align*}
 F_{\rm est, ent}   [M]  \ge     1-    O\left(  \frac 1 M\right) \, ,
 \end{align*} 
according to the central limit theorem.   Here, the estimation fidelity is defined as  
\[    F_{\rm est, ent}   [M]     =   \int  \d U \,  \int \d  \hat U  \,   p_M  (  \Phi_{\hat U}|   \Phi_U)   \,     \left|   \<  \Phi_{\hat U}  |  \Phi_U   \>  \right|^2  \,  , \]
where $p_M (  \Phi_{\hat U}|    \Phi_U)$ is the probability distribution resulting from the estimation  of a maximally entangled state with  $M$ nearly perfect copies.   
 Now,  we regard the estimation of the maximally entangled state  $|\Phi_U\>$   with $M$ copies as a particular strategy for the estimation of the unitary gate $U$ with $N$ copies,  meaning that we   have  
 \[  p_N  (\hat U  |  U)   \equiv   p_M     ( \Phi_{\hat U}|    \Phi_U   ) \,  \qquad  M  =    \Theta  (N^{2}/\log N ) \, .\]  
 Then, the estimation fidelity for the maximally entangled state can be easily converted into the gate fidelity for the corresponding gate, using the relation    \cite{horodecki-horodecki-1999-pra} 
\[F_{\rm  est, gate}   [  N]     = \frac{ ( d+1)  \,    F_{\rm est, ent}   [M]   - 1} d \,  . \] 
Since the state estimation fidelity converges to 1 as $1/ M$,   the gate estimation fidelity will also converge to 1 as  $1/M$.  Recalling that $M$ scales as $N^{2}/\log N$,   this proves Eq. (\ref{gateestimation}). 
  The error scaling   $\<  e\>    =  \log N/N^{2}$   beats the central limit scaling of classical statistics and is close to the optimal quantum  scaling $1/N^2$, which was derived in Refs. \cite{chiribella-dariano-2004-prl,bagan-baig-2004-pra,hayashi-2006-pla} for qubits and in Ref. \cite{kahn-2007-pra} for general $d$-dimensional systems.      The usefulness of our new derivation  is that the proof is much simpler than the full optimization of the estimation strategy.  
  
 It is interesting to mention that, for $d=2$, an alternative estimation strategy  achieving scaling $\log N^2/N^2$ was proposed by Rudolph and Grover \cite{rudolph-2003-prl}.   Their protocol is sequential and uses unentangled states to reach a quasi-Heisenberg scaling. It is an open question whether a similar protocol exists for $d>2$.    

\medskip  

\noindent {\bf  Universal supergeneration of pure states.}  
We now show that all  pure states can be supergenerated.  To this purpose, we parametrize the manifold of pure states as  $\{  |\psi_U\>      =   U \,  |0\> ~|~   U \in\grp {SU}  (d)\}$, where $|0\>$ is a fixed pure state. To achieve supergeneration, we use a classical strategy based on the estimation of the unknown gate $U$ and on the preparation of the state $|\psi_{\hat U}\>^{\otimes M}$  conditional on the estimate $\hat U$.   The fidelity of this strategy is given by 
\[  F_{\rm pure}  [  N\to M]  =    \int \d U  \, \int \d  \hat U  \,    p_N(\hat U | U)  \,     | \<\psi_{\hat U}    |\psi_U\>|^{2M}  \, , \]
where $p_N(\hat U|U)$ is the probability distribution resulting from gate estimation. 
 The above choice of strategy implies that we have the  bound 
\begin{align}\label{Fpure}
F_{\rm pure} [  N\to M]  \ge \left(  F_{\rm pure}  [N\to 1]\right)^M \, . 
\end{align}  
 Now, note that  the single-copy fidelity satisfies the relation  
\begin{align}
\nonumber F_{\rm pure} [N\to 1]  &  =      \int \d V \, \int \d U  \, \int \d  \hat U  \,    p_N(\hat U  V   | U V )  \,     | \<\psi_{\hat U  V}    |\psi_{U  V}\>|^{2}   \\
\nonumber  &  =      \int \d V \,  \int \d U  \, \int \d  \hat U  \, p_N(\hat U    | U )  \,     | \<\psi_{\hat U V}     |\psi_{  UV}\>|^{2}   \\   
\nonumber  &  =            \int \d V \, \int \d U  \, \int \d  \hat U  \,    p_N(\hat U    | U )  \,     | \<\psi_{V} |   \,   \hat U^\dag     U \,    |\psi_{V}\>|^{2}   \\   
 \nonumber &  =      \, \int \d U  \, \int \d  \hat U  \,    p_N(\hat U    | U )  \,      F_{\rm gate}  (\hat U ,  U)       \\   
\label{above} &  =  F_{\rm est, gate}  [N]  \, ,
\end{align}
where $F_{\rm est, gate}  [N]  $    is the gate estimation fidelity defined in Eq. (\ref{fgate}) and in the second equality we used the fact that the optimal gate estimation strategy is covariant \cite{book-holevo-probabilistic},~i.~e.~it satisfies the condition $  p_N(   \hat UV,UV)=  p_N (  \hat U , U)$ for every $\hat U, U, V  \in  \grp {SU}( d)$.  
 
 Combining Eq. (\ref{above})  with the bounds  (\ref{Fpure}) and (\ref{gateestimation}) we finally obtain  
\[  F_{\rm pure}  [N\to M]  \ge    1   - O  \left(      \frac {M  \,  \log N}  { N^{2}}\right)  \, ,\] 
meaning that arbitrary pure states can be supergenerated at every rate smaller than quadratic. Since the above protocol is based on estimation, the error vanishes only with a power law. It is an open question whether there exists a universal quantum supergeneration protocol whose error vanishes faster than any inverse polynomial.

 \medskip 

\noindent  {\bf  Equivalence between gate superreplication and gate estimation.}    As in the case of states, superreplication can be achieved by via gate estimation. The argument is the same used in the derivation of Eq. (\ref{bernoulli}): in short, one can show that every estimation strategy with fidelity scaling as $F_{\rm est, gate}[N]  = 1- O(   1/N^{\beta}) $  can be used to achieve gate replication with fidelity  
\begin{align}\label{classicalgate}  F_{  \rm gate}  [  N\to  M]   \ge   1  -   O  (  M/N^{\beta})  \,  , 
\end{align}
meaning that every replication rate smaller than $\beta$ can be achieved.    In particular, we know from Eq. (\ref{gateestimation}) that we can choose $\beta  =  2-  \epsilon$. Hence, every replication rate smaller than quadratic can be achieved via state estimation.   Note that, however, the replication error goes to zero only with a power law, while the quantum replication network has a much better scaling of the  error.  

The connection between gate superreplication and estimation can be used to prove the Heisenberg limit  \cite{giovannetti-lloyd-2004-science}.  
   The proof is by contradiction: 
   Suppose that one could estimate the parameters of a unitary gate with fidelity scaling as   $F_{\rm est, gate}[N]  = 1- O(   1/N^{2+\epsilon})$. Then, Eq. (\ref{classicalgate}) would imply  that one can simulate $M  =    O\left(  N^{2 +\epsilon/2}\right)$ copies of the gate, in contradiction with Theorem \ref{theo:strong4}.  In summary, the quadratic scaling of the Heisenberg limit can be derived solely from considerations about the optimal rates of gate superreplication.

\medskip
\noindent{\bf Asymptotic no-cloning theorem for quantum gates in the worst case scenario.}  We have seen that gate superreplication can be achieved on all input states except for a vanishingly small fraction.   A natural question is whether one can find a superreplication protocol that works on \emph{all} input states.    The problem can be addressed by evaluating the worst-case fidelity
\begin{align}\label{F-gate-worst}
F_{\rm gate, worst}  [N\to M]=\min_{  |\Psi\>\in\spc{H}^{\otimes M},   \|   |\Psi\>   \|  = 1}    \quad   \int \d U    \,  \<\Psi|(U^\dag)^{\otimes M}\map{C}^{(N)}_U\left(|\Psi\>\<\Psi|\right)U^{\otimes M}|\Psi\>    \, ,
\end{align}
where $\map{C}^{(N)}_U $ is the channel implemented by the replication network. 
In this setting, we say that a gate replication process is \emph{reliable in the worst case} iff
\[  \lim_{N\to \infty}  F_{\rm gate, worst}  [N\to M(N)]   =  1  \]
and we say that the replication rate $\alpha$ is \emph{achievable in the worst case} iff there exists a replication protocol that has that rate and is reliable in the worst case.  The  supremum over all achievable rates is determined by the following 
 \begin{theo}[Asymptotic no-cloning theorem for quantum gates]
  For finite dimensional quantum systems,  no physical  process can reliably  replicate   phase gates  at rate $\alpha  \ge 1$  in the worst case scenario.  
\end{theo}
Here we give a heuristic proof, based on optimal phase estimation.  When an unknown phase gate is used $N$ times,  the optimal estimation strategy has mean square error $c/N^2$, for some suitable constant $c>0$ \cite{buzek-derka-1999-prl,berry-wiseman-2000-prl}.   The bound holds both for deterministic and  probabilistic strategies \cite{chiribella-dariano-2008-prl-memory} and the fact that the optimal estimation strategy can be achieved deterministically plays a crucial role in our argument.        Now, suppose that one can  produce $M$ reliable replicas, up to an error scaling as $1/M^\beta$, for some $\beta  >  2$.   In this case, one could use the replicas for phase estimation, reducing the error to 
\[\<e\>   =  \frac c {M^2}  +  O\left ( \frac 1 {M^\beta}  \right) \, , \]
this result following the fact that the error of a deterministic estimation strategy is a continuous  function of the  input state.  
Since the above  strategy cannot be better than the optimal strategy, we obtain the inequality  
\begin{align}\label{scaling}  \frac c {N^2}  \le \frac   c {M^2}     +  O\left( \frac1{ M^\beta} \right)  \,,
\end{align}  which implies that $M$ can grow at most as $  M  =  N  +   \Theta ( N^\alpha)$ with $\alpha <1$.    In addition, by Taylor-expanding the r.h.s. of Eq. (\ref{scaling}) we obtain the inequality $\beta  < 3-\alpha$, meaning that the error can vanish at most  as $1/N^3$.    In summary, superreplication and exponentially vanishing errors are forbidden in the worst case scenario.

\iffalse
 In infinite dimensions, the above argument does not work, as there  is no limit to the precision in the estimation of a phase gate, unless one poses limits input states used to probe the gate. A possible generalization of our theorem to infinite dimensions would involve a state estimation argument using input states with   bounded energy, or with bounded expectation value of  some other distinguished observable.   
\fi
 
%State generation is a task that is less demanding than state cloning:indeed, processing the state $|\psi\>$ is not equivalent to processing the gate $U$. The state $|\psi\>$ can be easily generated by applying $U$ to $|\psi_0\>$, but the inverse is prohibited by the no-programming theorem \cite{nielsen-chuang-1997-prl}.  The first thing to see is that, if the quadratic number of replicas could be achieved, it would help to build a searching algorithm which outputs all the desired data in a size-$M$ database $\{|1\>,|2\>,\dots,|M\>\}$, querying the oracle only $O(\sqrt{M})$ times. Indeed, the algorithm would need only to superrelicate the oracles and then apply the network to the product state $|1\>\otimes\cdots\otimes|M\>$. Measuring the output state would reveal all the desired entries. The algorithm, however, would not require the information about the amount of desired data, and would thus contradict the optimality of Grover $\pi/3$ algorithm, which needs $O(M)$ queries of the oracle.

\section{Conclusion and outlook}\label{sec:conclusion}
In this paper we reviewed the phenomenon of superreplication of quantum states and gates, emphasizing the applications and connections with other tasks in quantum information.   In particular, we clarified the relation between  superreplication and the precision limits of quantum metrology, showing that  \emph{i)} estimation can be used to achieve superreplication with error vanishing as $O(M/N^2)$, probabilistically in the case of states and deterministically in the case of gates, \emph{ii)} gate superreplication allows for a simpler proof of the quadratic precision enhancement in the estimation of an unknown gate, \emph{iii)}   the optimality of the  Heisenberg scaling for the estimation of unitary gates can be derived from the ultimate limit to the rate of superreplication.     
In addition, we showed that $N$ uses of a completely unknown gate are sufficient to generate $O(N^2 )$ approximate copies of the corresponding pure state, with an error that vanishes in the large $N$ limit.   

Among the future research directions, an important one is the study of replication processes for mixed states and noisy channels. We expect that also in this case there is an equivalence between  replication  and estimation.   In the presence of noise, it is well known that the Heisenberg scaling is often inhibited, leading to different sub-Heisenberg scalings \cite{dorner-demkowicz-2009-prl,escher-dematos-2011-natphys,demkowicz-koodyski-2012-natcomm,chaves-brask-2013-prl,demkowicz-maccone-2014-prl}. We then expect that the replication rates will also have intermediate values, ranging between $\alpha=1$ and $\alpha=2$ depending on the type of noise.    The study of superreplication in the noisy case is also expected to shed light on the optimal  strategies for the estimation of noisy channels, whose performances are sometimes hard to characterize analytically.  
Another interesting research direction  is the study of replication processes that are subject to constraints,~e.~g.~on the ability to perform joint  operations on composite systems. Recently,  Kumagai and Hayashi \cite{kumagai-hayashi-2013-arxiv} investigated the problem of cloning bipartite quantum states using only \emph{local operations} and classical communication.  
When the state to be cloned is perfectly known, they showed that the number of extra copies scales as $\sqrt N$.   Note the contrast with the situation where the LOCC restriction is not imposed, in which case one can produce $\Theta (N^{\delta})$ extra copies for every $\delta <1$.    Besides the constraints due to the locality of the operations, other  physical constraints can arise from conservation laws, such as energy and angular momentum conservation  \cite{ozawa-2002-prl, banacloche-ozawa-2005-job, ahmadi-jennings-rudolph-2013-njp,marvian-spekkens-2013-njp, marvian-spekkens-2014-natcomm}. The search for energy-preserving cloners has been considered in our earlier work \cite{yang-chiribella-2015-arxiv}, where we identified the optimal operations for the transformation of pure states.     In this scenario, it is interesting to examine  how the replication rates are affected by the presence of limited resources, or, conversely, what are the resources needed to achieve a desired replication rate.   

\section*{Acknowledgements} This work is supported  by the Foundational Questions Institute (FQXi-RFP3-1325),  the National Natural Science Foundation of China (11450110096, 11350110207), and the 1000 Youth Fellowship Program of China.
 % Moreover, we explored the possibility to  reduce the size of the interactions needed to achieve  large-scale quantum cloning.  We considered protocols where the   $N$ input states/gates are divided into smaller groups of size $M/N$, which interact jointly with   $O(M^2/N^2)$ blank copies.   

\bibliography{ref}
\bibliographystyle{unsrt}

\end{document}

%% file: myQcircuit.tex
\usepackage[matrix,frame,arrow]{xy}
\usepackage{amsmath}

\newcommand{\qw}[1][-1]{\ar @{-} [0,#1]}
\newcommand{\multigate}[2]{*+<1em,.9em>{\hphantom{#2}} \qw \POS[0,0].[#1,0];p !C *{#2},p \save+LU;+RU **\dir{-}\restore\save+RU;+RD **\dir{-}\restore\save+RD;+LD **\dir{-}\restore\save+LD;+LU **\dir{-}\restore}

\newcommand{\Qcircuit}[1][0em]{\xymatrix @*[o] @*=<#1>}  %tentativo disperato
 \renewcommand{\Qcircuit}[1][0em]{\xymatrix @*=<#1>}

\newcommand{\pureghost}[1]{*+<1em,.9em>{\hphantom{#1}}}
\newcommand{\poloFantasmaCn}[1]{{{}^{#1}_{\phantom{#1}}}}